\documentclass[aps,nofootinbib,showkeys,preprint]{revtex4}

\usepackage{epsf,epsfig,subfigure,graphicx,amsmath,amssymb}
\usepackage{color}

\begin{document}

\title{\bf   Boosted Dark Matter Signals Uplifted with Self-Interaction  }

\author{
Kyoungchul Kong$^{(a)}$\footnote{email: kckong@ku.edu }, Gopolang Mohlabeng$^{(a)}$\footnote{email: mohlabeng319@gmail.com}
and Jong-Chul Park$^{(a, b)}$\footnote{email: log1079@gmail.com}
}
\affiliation{$^{(a)}$ Department of Physics and Astronomy, University of Kansas, Lawrence, KS 66045, USA\\
$^{(b)}$ Department of Physics, Sungkyunkwan University, Suwon 440-746, Republic of Korea
}

\begin{abstract}
We explore detection prospects of a non-standard dark sector in the context of boosted dark matter.
We focus on a scenario with two dark matter particles of a large mass difference,
where the heavier candidate is secluded and interacts with the standard model particles only at loops,
escaping existing direct and indirect detection bounds.
Yet its pair annihilation in the galactic center or in the Sun may produce boosted stable particles,
which could be detected as visible Cherenkov light in large volume neutrino detectors.
In such models with multiple candidates,
self-interaction of dark matter particles is naturally utilized in the {\it assisted freeze-out} mechanism and is
corroborated by various cosmological studies such as N-body simulations of structure formation, observations of dwarf galaxies, and the small scale problem.
We show that self-interaction of the secluded (heavier) dark matter greatly enhances the capture rate in the Sun
and results in promising signals at current and future experiments.
We perform a detailed analysis of the boosted dark matter events for Super-Kamiokande, Hyper-Kamiokande and PINGU,
including notable effects such as evaporation due to self-interaction and energy loss in the Sun.

\end{abstract}


\maketitle

\section{Introduction}

Dark matter (DM) is one of the most profound mysteries in particle physics and cosmology.
Recent observations show that 25\% of our universe is made up of dark matter,
yet we know very little about its nature and properties.
Especially its microscopic nature such as its stabilizing mechanism, spin and mass,
necessitates a balanced program based on various dark matter searches \cite{Arrenberg:2013rzp}.

Among a myriad of possibilities, scenarios with multiple dark matter particles are well motivated and
their implications have been studied at different scales from the large in cosmology to the small at the Large Hadron Collider (LHC) at CERN \cite{Konar:2009qr}.
Several issues have been especially investigated on the cosmological side in the context of multiple dark matter candidates.
While N-body simulations of structure formation based on cold dark matter (CDM) present a steep cusp density profile \cite{Moore:1994yx},
observations of dwarf galaxies indicate a cored density profile rather than a cusped one \cite{Walker:2011zu} (so-called the ``core vs cusp problem'').
Simulations also predict that CDM evolves to very dense subhalos of Milky Way type galaxies, which can not host the brightest satellites, but it would be hard to miss the observation of these substructures (known as the ``too big to fail problem'') \cite{BoylanKolchin:2011de}.
Warm dark matter has been proposed as a solution to the small scale conflict between the observations and the simulations with CDM,
since it is expected to develop shallower density profiles at a small scale and would avoid unreasonably dense subhalos \cite{Lovell:2013ola}.

Self-interacting DM (SIDM) has been suggested as another interesting solution to those small scale problems~\cite{Spergel:1999mh}.
Cosmological simulations with SIDM \cite{Rocha:2012jg} show that SIDM with the ratio of the DM self-interaction cross section to the DM mass $\sigma_{\chi\chi}/m_\chi \sim \mathcal{O}(0.1-1\,{\rm cm}^2/{\rm g})$
can reconcile the inconsistency between simulations and observations at a small scale, while it does not modify the CDM behavior at a large scale.
Analysis of the matter distribution of the Bullet Cluster \cite{Randall:2007ph} provides the most robust constraint on SIDM, $\sigma_{\chi\chi}/m_\chi < 1.25\,{\rm cm}^2/{\rm g}$.
Another analysis based on the kinematics of dwarf spheroidals \cite{Zavala:2012us} shows that SIDM resolves the small scale conflicts of CDM only when $\sigma_{\chi\chi}/m_\chi  \gtrsim 0.1\,{\rm cm}^2/{\rm g}$.

In this paper, we investigate detection prospects of two-component dark matter at large volume neutrino detectors.
We focus on a scenario with a relatively large mass gap between the two components,
where the heavier candidate interacts with the standard model (SM) particles only at loops.
Its sister (the light one) is assumed to have interactions with both the heavier counterpart and the standard model particles.
If the heavier dark matter is dominant in our current universe,
the dark sector with such candidates is secluded and all current direct and indirect bounds are evaded.
Although the light dark matter particles are subdominant,
they may be produced via the annihilation of the heavy sisters with a large boost due to the large mass difference.
A boosted DM arises in various multi-component DM scenarios such as semi-annihilation $\psi_i\psi_j \to \psi_k\phi$~\cite{D'Eramo:2010ep, Belanger:2012vp}, assisted freeze-out $\psi_i\psi_i \rightarrow \psi_j\psi_j$~\cite{Belanger:2011ww}, and decay $\psi_i \rightarrow \psi_j + \phi$.
Recently a possibility of detecting a boosted dark matter particle in large volume neutrino telescopes has been examined \cite{Huang:2013xfa, Agashe:2014yua, Berger:2014sqa}.
In Ref. \cite{Agashe:2014yua}, the heavier DM annihilates in the center of the galaxy, and its pair annihilation products travel to the Earth and
leave Cherenkov light in the detector via a neutral current-like interaction, which points toward the galactic center (GC).
Detection of boosted dark matter from the Sun has been studied in Ref. \cite{Berger:2014sqa}, where
a search for proton tracks pointing toward the Sun is proposed in a different model.

We explore detection prospects of boosted dark matter from the Sun in the presence of self-interaction of the heavier component,
which is well motivated by various cosmological studies as mentioned earlier.
We include important effects that are neglected in literature such as evaporation of the dark matter and energy loss during traveling from the core to the surface of the Sun.
As a concrete example, we consider a model that was studied in Ref. \cite{Agashe:2014yua}, which is revisited in Section \ref{sec:model}.
A detailed calculation of the boosted dark matter flux is outlined in Section \ref{sec:flux}, and detection prospects in Section \ref{sec:detection}.
We focus on the discovery potential at Super-Kamiokande(Super-K), Hyper-Kamiokande(Hyper-K), and PINGU.

\section{Boosted dark matter in assisted freeze-out}
\label{sec:model}

In this section, we present an explicit example of a model with two-component DM in order to discuss detection prospects of boosted DM from the Sun.
We choose the model studied in Ref.~\cite{Agashe:2014yua} based on the assisted freeze-out mechanism~\cite{Belanger:2011ww}.
Additionally we introduce DM self-interaction preferred by cosmological simulations and observations for the heavier constituent of the two DM components.
We only briefly summarize the key points of our bench mark model and refer to Ref.~\cite{Agashe:2014yua} for details on the model.

\subsection{Basic set-up}

\begin{figure}
\begin{center}
\hspace{0.3cm}
\includegraphics[width=0.32\linewidth]{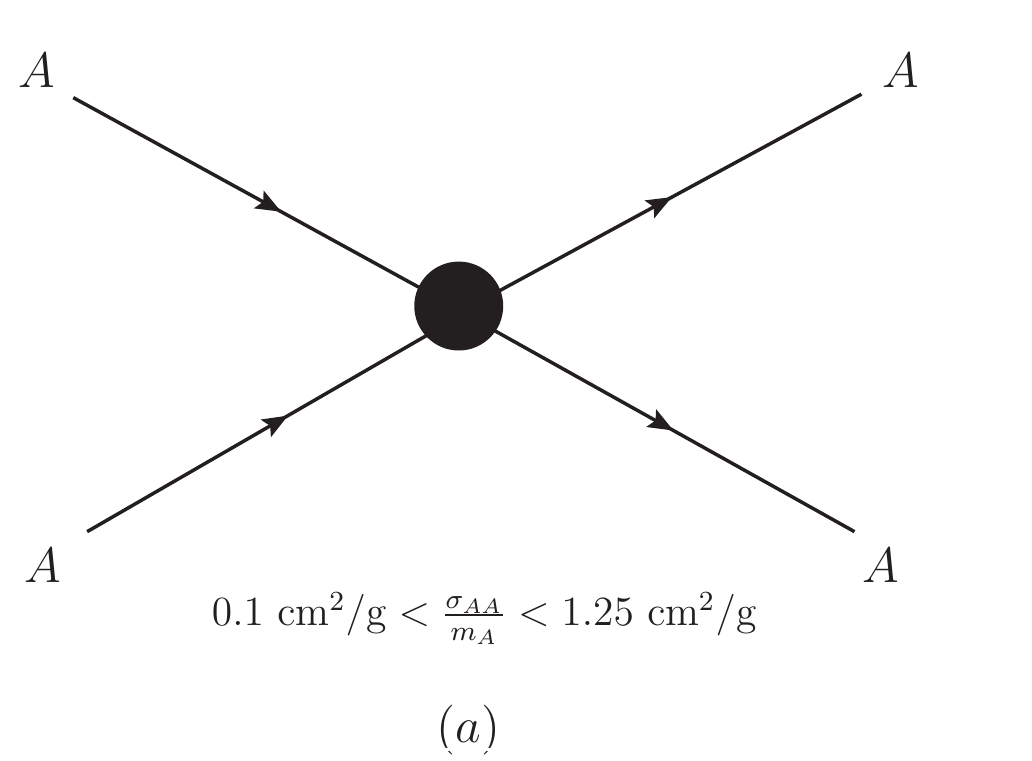}
\hspace*{0.01cm}
\includegraphics[width=0.32\linewidth]{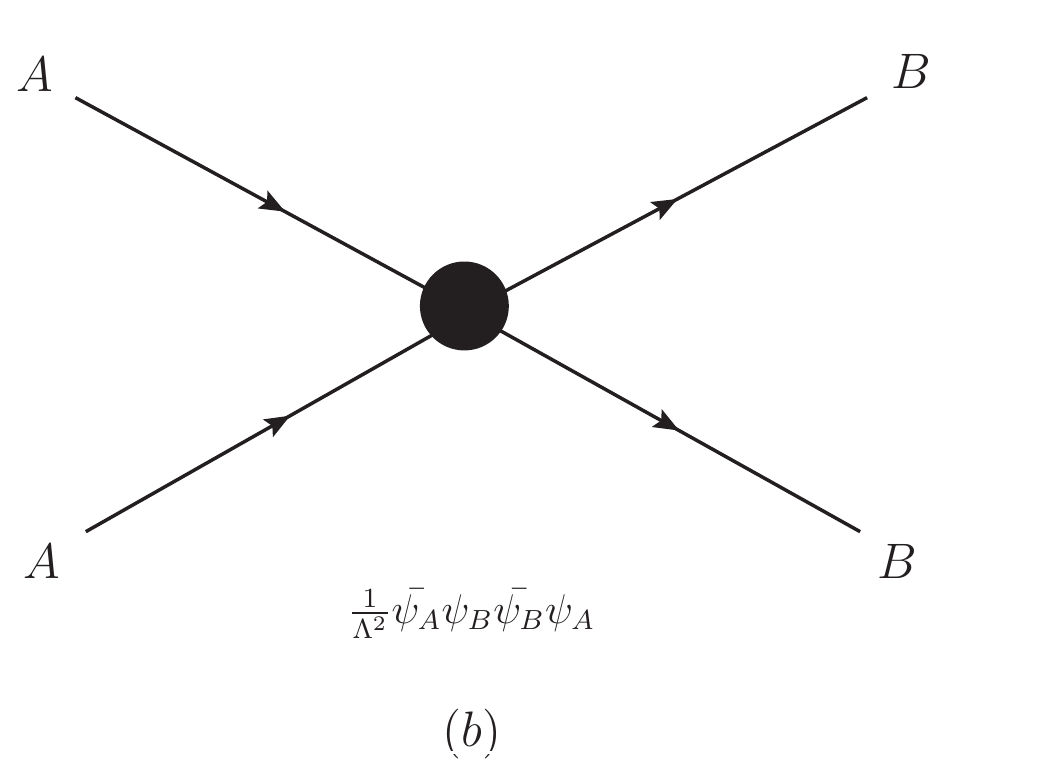}
\hspace*{0.01cm}
\includegraphics[width=0.30\linewidth]{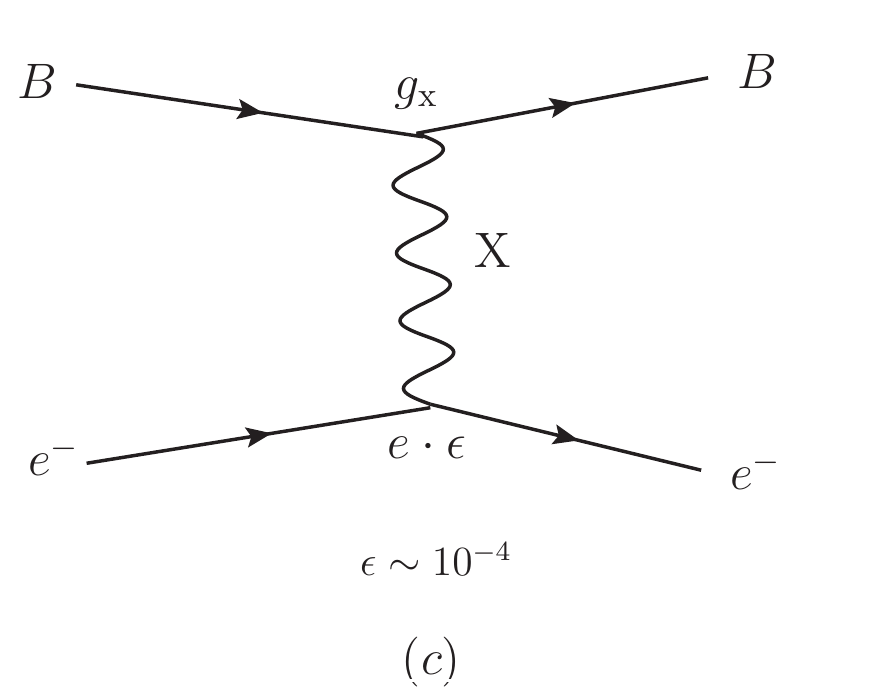}
\end{center}
\vspace*{-0.7cm}
\caption{
Diagrams for (a) self-interaction of the heavier DM $\psi_A$, (b) production of the boosted DM $\psi_B$ from the annihilation of $\psi_A$, and (c) elastic scattering of $\psi_B$ off an electron.
}
\label{Diagrams}
\end{figure}

We consider the case where $\psi_A$ and $\psi_B$ are two stable DM candidate particles with masses $m_A > m_B$.
This can be achieved with separate symmetries, for example,  ${\rm U(1)}'\otimes{\rm U(1)}''$~\cite{Belanger:2011ww} or $Z_2\otimes Z_2'$~\cite{Agashe:2014yua}.
We assume that two DM species, $\psi_A$ and $\psi_B$ interact via a contact operator,
\begin{eqnarray}\label{ContactOp}
\mathcal{L}_{AB} = \frac{1}{\Lambda^2} \overline{\psi}_A \psi_B \overline{\psi}_B \psi_A \, ,
\end{eqnarray}
and that $\psi_A$ can only annihilate into $\psi_B$ and not directly into SM particles.
Moreover, the heavier component $\psi_A$ is the dominant DM constituent in the universe.
The boosted DM $\psi_B$ is currently produced via the contact interaction Eq.~(\ref{ContactOp}).
We additionally allow a self-interaction for $\psi_A$ in the range of
$0.1\,{\rm cm}^2/{\rm g} < \sigma_{AA}/m_A < 1.25\,{\rm cm}^2/{\rm g}$ (Figure \ref{Diagrams}(a)),
favored by simulations and observations~\cite{Spergel:1999mh, Rocha:2012jg, Randall:2007ph, Zavala:2012us}.

%
%

The particle $\psi_B$ is charged under a hidden ${\rm U(1)}_X$ gauge symmetry, with a charge $Q_X^B=1$ for simplicity, which is spontaneously broken leading to the gauge boson mass $m_X$.
In addition, a mass hierarchy, $m_A > m_B > m_X$ is assumed.
The gauge coupling of ${\rm U(1)}_X$, $g_X$ will be taken to be large enough, e.g. $g_X=0.5$, so that the thermal relic density of $\psi_B$ is small due to the large annihilation cross section of the process $\psi_B \overline{\psi}_B \to XX$.
We assume that the DM sector couples to the SM sector only through a kinetic mixing between ${\rm U(1)}_X$ and ${\rm U(1)}_{\rm EM}$ (originally ${\rm U(1)}_Y$)~\cite{Okun:1982xi, Chun:2010ve},\footnote{One can find a general and detailed analysis on a hidden sector DM and the kinetic mixing in Ref.~\cite{Chun:2010ve}.}
\begin{eqnarray}\label{mixing}
\mathcal{L} \supset -\frac{1}{2}\, \sin\epsilon\, X_{\mu\nu}F^{\mu\nu}\,.
\end{eqnarray}
Thus, $\psi_B$ can scatter off SM particles via a $t-$channel $X$ boson exchange.

This model can be described by a set of seven parameters:
\begin{eqnarray}
\{m_A, m_B, m_X, \Lambda, g_X, \epsilon, \sigma_{AA}\}\,,
\end{eqnarray}
where $\Lambda$ will be appropriately taken in our analysis to obtain the required DM relic density, $\Omega_A\simeq \Omega_{\rm DM} \approx 0.2$, as done in Ref.~\cite{Agashe:2014yua}.
In all the interactions between DM and SM particles, $g_X$ and $\epsilon$ always appear as a simple combination, $(g_X \cdot \epsilon)$.
As a result, our analysis will mainly rely on five parameters, $\{ m_A, m_B, m_X, g_X \cdot \epsilon, \sigma_{AA} \}$.
For easier comparison, we choose the same benchmark scenario as in Ref.~\cite{Agashe:2014yua},
except for $\epsilon$,
\begin{eqnarray}\label{benchmark}
m_A = 20\,{\rm GeV},~ m_B = 0.2\,{\rm GeV},~ m_X = 20\,{\rm MeV},~ g_X = 0.5,~ {\rm and}~ \epsilon = 10^{-4}\,.
\end{eqnarray}
However, we choose $\epsilon=10^{-4}$, instead of $10^{-3}$ chosen as a reference value in Ref.~\cite{Agashe:2014yua}, for boosted $\psi_B$ to avoid too much energy loss during traversing the Sun as explained in Section~\ref{EnergyLoss}.
$\epsilon=10^{-4}$ is well consistent with current limits on a hidden $X$ gauge boson (or a dark photon), $\epsilon \lesssim \mathcal{O}(10^{-3})$ for $m_X \gtrsim$ 10 MeV~\cite{Essig:2013lka}.

\subsection{Relic abundance and scattering cross sections}\label{XSection}

A set of coupled Boltzmann equations describes the evolution of the relic density of two DM particles, $\psi_A$ and $\psi_B$, in the assisted freeze-out mechanism~\cite{Belanger:2011ww, Agashe:2014yua, Bandyopadhyay:2011qm}.\footnote{
See Ref.~\cite{Belanger:2011ww} for a numerical analysis and Ref.~\cite{Agashe:2014yua} for more details on analytic estimates.}
The annihilation process $\psi_A \overline{\psi}_A \to \psi_B \overline{\psi}_B$ (Figure~\ref{Diagrams}(b)) determines the thermal relic abundance of $\psi_A$ as well as the production rate of boosted $\psi_B$ in the current universe.
The annihilation cross section for the process is obtained as
\begin{eqnarray}
\langle \sigma_{A\overline{A} \to B\overline{B}}\, v \rangle \simeq
\frac{1}{8\pi \Lambda^4}\, (m_A + m_B)^2\, \sqrt{1-\frac{m_B^2}{m_A^2}}\, +\, \mathcal{O}(v^2)
\end{eqnarray}
from the contact operator in Eq.~(\ref{ContactOp}).
In the limit $\langle \sigma_{B\overline{B} \to XX}\, v \rangle \gg \langle \sigma_{A\overline{A} \to B\overline{B}}\, v \rangle$, the relic abundance of $\psi_A$ is given by~\cite{Agashe:2014yua}
\begin{eqnarray}
\Omega_A \simeq 0.2\, \left( \frac{5\times 10^{-26}\, {\rm cm}^3/{\rm  s}}{\langle \sigma_{A\overline{A} \to B\overline{B}}\, v \rangle}  \right)\,.
\end{eqnarray}
Indeed, $\langle \sigma_{B\overline{B} \to XX}\, v \rangle \gg \langle \sigma_{A\overline{A} \to B\overline{B}}\, v \rangle$ corresponds to the case that we are interested in, and thus the abundance of $\psi_A$ dominates over that of $\psi_B$.
In our numerical analysis, we will set  $\langle \sigma_{A\overline{A} \to B\overline{B}}\, v \rangle \simeq 5\times 10^{-26}\, {\rm cm}^3/{\rm  s}$.

The lighter component $\psi_B$ can scatter off SM particles via a $t-$channel $X$ boson exchange through the kinetic mixing as shown in Eq.~(\ref{mixing}).
However, we cannot detect signals from scattering off nuclei by the thermal relic $\psi_B$ in dark matter direct detection experiments due to its tiny abundance, e.g., $\Omega_B \thickapprox \mathcal{O}(10^{-7}-10^{-6})$ for the benchmark scenario in Eq.~(\ref{benchmark}).
As shown in Ref.~\cite{Agashe:2014yua}, the boosted $\psi_B$ from the process $\psi_A \overline{\psi}_A \to \psi_B \overline{\psi}_B$ might be detected at a large volume neutrino detector through its elastic scattering off electrons, $\psi_B e^- \to \psi_B e^-$ (Figure \ref{Diagrams}(c)).
The minimum detectable scattered electron energy is set by the threshold energy of each experiment, $E_e^{\rm min} = E_e^{\rm th}$, and
the maximum energy is given by
\begin{eqnarray}\label{Emax}
E_e^{\rm max} = m_e\, \frac{(E_B + m_e)^2 + E_B^2 - m_B^2}{(E_B + m_e)^2 - E_B^2 + m_B^2}\,,
\end{eqnarray}
where $E_B$ is the energy of the boosted $\psi_B$ before its collision with a target electron.
The differential cross section for the process $\psi_B e^- \to \psi_B e^-$ is given by
\begin{eqnarray}\label{dsigBe}
\frac{d\sigma_{B e^- \to B e^-}}{dt} = \frac{1}{8\pi}\, \frac{(e\epsilon g_X)^2}{(t-m_X^2)^2}\, \frac{8 E_B^2 m_e^2 + t(t + 2s)}{s^2 + m_e^4 + m_B^4 - 2 s m_e^2 -2 s m_B^2 - 2 m_e^2 m_B^2}\,,
\end{eqnarray}
where $s = m_B^2 + m_e^2 + 2E_B m_e$ and $t = 2 m_e (m_e - E_e)$.

The heavier DM $\psi_A$ can interact with the SM sector via a $\psi_B$ loop even though $\psi_A$ has no direct coupling to the SM sector.
The $\psi_A$-nucleon scattering cross section is
\begin{eqnarray}\label{sigAn}
\sigma_{A-{\rm nucleon}} = \frac{\mu_{A-p}^2 (Z \epsilon e)^2}{\pi A^2}\, \frac{t^2}{(t - m_X^2)^2}\, \left[ \frac{g_X}{48 \pi^2}\, \frac{\log (m_B^2/(\lambda\Lambda)^2)}{\Lambda^2} \right]^2\,,
\end{eqnarray}
where $\mu_{A-p}$ is the $\psi_A$-nucleon reduced mass, $A$ and $Z$ denote the atomic mass and the proton number of a target nucleus, $t=-2 m_N E_R$ with the nucleus mass $m_N$ and the nucleus recoil energy $E_R$, $\lambda$ is a hidden sector Yukawa coupling of order unity, and $\Lambda$ is determined by $\langle \sigma_{A\overline{A} \to B\overline{B}}\, v \rangle \simeq 5\times 10^{-26}\, {\rm cm}^3/{\rm  s}$~\cite{Agashe:2014yua}.
The cross section $\sigma_{A-{\rm nucleon}}$ is suppressed by the small $\epsilon$ parameter and one-loop factor, and thus the direct detection of the relic $\psi_A$ is almost impossible even in a future DM direct detection experiment, e.g. XENON1T.
However, even this small cross section will contribute to the accumulation of $\psi_A$ in the Sun.

\section{Boosted dark matter flux from the Sun}
\label{sec:flux}

In this section, we briefly review the evolution of the DM ($\psi_A$) number in the Sun, and calculate the boosted DM ($\psi_B$) flux from the annihilation of the heavier DM component ($\psi_A$).
The DM capture in the Sun via the collisions between DM and nuclei was examined in Refs.~\cite{Steigman:1997vs, Griest:1986yu}.
Subsequent studies discussed several important effects such as evaporation for a relatively light DM ($m_{\rm DM} \lesssim$ 3-5 GeV) \cite{Gould:1987ju,Busoni:2013kaa} and enhancement of the DM accumulation due to self-interaction \cite{Zentner:2009is, Albuquerque:2013xna}.
Such a DM self-interaction has been proposed to alleviate the small scale structure problems of simulations with collisionless CDM~\cite{Spergel:1999mh}.
It has been shown that the self-interaction also participates in the evaporation process reducing the DM number \cite{Chen:2014oaa}.

\subsection{Evolution of dark matter in the Sun}

The time evolution of the DM number $N_\chi$ in the Sun is described by the following differential equation~\cite{Chen:2014oaa}
\begin{eqnarray}\label{N-evolution}
\frac{dN_\chi}{dt} = C_c + (C_s-C_e)N_\chi - (C_a+C_{se})N_\chi^2\,,
\end{eqnarray}
where $C_c$ is the DM capture rate by the Sun, $C_s$ is the DM self-capture rate, $C_{e}$ is the DM evaporation rate due to DM-nuclei interactions, $C_{a}$ is the DM annihilation rate, and $C_{se}$ is the evaporation rate due to DM self-interaction.
In our analysis, we assume that the DM and nuclei inside the Sun follow a thermal distribution, and thus use numerical data on the solar model such as mass density $\rho(r)$, temperature $T(r)$, and mass fraction of the atom $i$, $X_i(r)$ inside the Sun given in Ref.~\cite{SolarModelFile}.

If a DM particle interacts with nuclei, it loses its kinetic energy during traveling inside the Sun.
The DM particle is gravitationally captured when its final velocity after the collision with nuclei is smaller than the escape velocity $v_{\rm esc}(r)$ from the Sun.
The number of DM particles in the Sun increases through this capture process.
The DM capture rate in the Sun $C_{c}$ has been investigated in Refs.~\cite{Kappl:2011kz, Busoni:2013kaa, Baratella:2013fya}.
In our study, we use the numerical results from Ref.~\cite{Busoni:2013kaa} ($m_\chi \lesssim$ 10 GeV) and Ref.~\cite{Baratella:2013fya} ($m_\chi \gtrsim$ 10 GeV).
For more details on the exact calculation, see Refs.~\cite{Kappl:2011kz, Busoni:2013kaa}.

The coefficient $C_{a}$ describes the annihilation of two DM particles trapped inside the Sun, which has been well studied in Refs.~\cite{Steigman:1997vs, Griest:1986yu}.
Based on the exact numerical calculation, Refs.~\cite{Busoni:2013kaa, Kappl:2011kz} provided fitting functions: the former is valid in the range 0.1 GeV $\lesssim m_\chi \lesssim$ 10 GeV and the latter for $m_\chi \gtrsim$ a few GeV.
We adopt the fitting functions from Ref.~\cite{Busoni:2013kaa} ($m_\chi \lesssim$ 10 GeV) and Refs.~\cite{Kappl:2011kz} ($m_\chi \gtrsim$ 10 GeV).

A captured DM particle could scatter off energetic nuclei and escape from the Sun when its velocity after the scattering is larger than the local escape velocity $v_{\rm esc}(r)$, which is generally called the evaporation process~\cite{Griest:1986yu, Gould:1987ju}.
The basic idea of evaporation is the same as capture.
The main difference is whether the final velocity is smaller (for capture) or larger (for evaporation) than the escape velocity $v_{\rm esc}(r)$.
The evaporation rate $C_{e}$ is effective only for a low DM mass, $m_\chi \lesssim$ 5 GeV and completely negligible for heavier DM masses.
For the evaporation rate, the fitting functions to the numerical results given in Ref.~\cite{Busoni:2013kaa} are used in our analysis.
For more details on the calculation of $C_{e}$, see Refs.~\cite{Gould:1987ju, Busoni:2013kaa}.

Self-interactions of DM will also affect its capture and evaporation processes inside the Sun.
The $C_{s}$ is the self-capture rate by scattering off other DM particles that have already been trapped within the Sun.
In this DM-DM scattering, a target DM particle that obtains too much kinetic energy will be ejected from the Sun, which results in no net accumulation of DM particles unlike the capture by collision with nuclei.
However, the escape velocity from the interior of the Sun is at least two times larger than the typical velocity of a galactic DM particle.
Thus, the ejection of a target DM particle via the DM-DM collision results in a tiny correction to the typical DM self-capture rate in the Sun \cite{Zentner:2009is}
\begin{eqnarray}\label{SelfCapture}
C_s = \sqrt{\frac{3}{2}}\, n_\chi \sigma_{\chi\chi} v_{\rm esc}(R_\odot)\,
\frac{v_{\rm esc}(R_\odot)}{\overline{v}}\, \langle \widehat{\phi}_{\chi}  \rangle\, \frac{{\rm erf}(\eta)}{\eta}\,,
\end{eqnarray}
where $n_\chi$ is the local number density of galactic DM, $\sigma_{\chi\chi}$ is the self-elastic scattering cross section of DM, $v_{\rm esc}(R_\odot)$ is the escape velocity at the surface of the Sun,
$\langle \widehat{\phi}_{\chi} \rangle$ is a dimensionless average solar potential experienced by the captured DM within the Sun, and $\eta^2 = 3(v_\odot/\overline{v})^2/2$ is a dimensionless variable with the velocity of the Sun $v_\odot = 220 {~\rm km}/{\rm s}$ and the local velocity dispersion of DM $\overline{v} = 270 {~\rm km}/{\rm s}$.
$\langle \widehat{\phi}_{\chi} \rangle \simeq 5.1$~\cite{Gould:1991hx} is generally used in the calculation of $C_s$, which however deviates from the commonly used value for smaller DM masses, $m_\chi \lesssim$ 10 GeV.
Thus, we numerically calculate $\langle \widehat{\phi}_{\chi} \rangle$ for our analysis.
The full expression of the self-capture rate including the small ejection effect of the target DM particle is given in the Appendix of Ref. \cite{Zentner:2009is}.

The last coefficient is the self-interaction induced evaporation rate $C_{se}$.
A DM particle captured in the Sun can scatter off another captured DM particle through their self-interaction, which leads to evaporation when one of two colliding DM particles has velocity greater than the escape velocity $v_{\rm esc}(r)$ after the collision.
The authors of Ref.~\cite{Chen:2014oaa} recently investigated the self-interaction induced evaporation and provided details of the derivation of $C_{se}$ in the Appendix of their paper.
We numerically calculate $C_{se}$ based on the analytic expression given in the Appendix of Ref.~\cite{Chen:2014oaa}.
We assume that the DM temperature is in thermal equilibrium with the solar temperature following Ref.~\cite{Chen:2014oaa}.
Thus, we use the solar temperature $T$ as the DM temperature $T_\chi$ in our calculation.

\subsection{Accumulated dark matter number and annihilation rate}

With the initial condition $N_\chi(0)=0$, the solution to the DM evolution equation, Eq.~(\ref{N-evolution}) is given by~\cite{Chen:2014oaa}
\begin{eqnarray}\label{N-final}
N_\chi(t) = \frac{C_c\, \tanh(t/\tau_{\rm eq})}{\tau_{\rm eq}^{-1} - (C_s - C_e) \tanh(t/\tau_{\rm eq})/2}
\end{eqnarray}
with
\begin{eqnarray}\label{Equi-time}
\tau_{\rm eq} = \frac{1}{\sqrt{C_c(C_a + C_{se}) + (C_s - C_e)^2/4}}\,,
\end{eqnarray}
where the $\tau_{\rm eq}$ is the time-scale required for the DM number $N_\chi(t)$ in the Sun to reach equilibrium between accumulation by $C_c$ and $C_s$ and dissipation by $C_a, C_e$, and $C_{se}$.
Then, the DM annihilation rate inside the Sun is simply given by
\begin{eqnarray}\label{annihilation}
\Gamma_A^\chi = \frac{C_a}{2}\, N_\chi^2\,.
\end{eqnarray}
For the age of the Sun $t = t_\odot \simeq 4.6 \times 10^9$ year, we obtain the currently accumulated number and annihilation rate of DM in the Sun.
When the equilibrium state is attained, i.e., $t\gtrsim\tau_{\rm eq}$, $N_\chi$ and $\Gamma_A^\chi$ can be simplified as
\begin{eqnarray}\label{Neq}
N_\chi^{\rm eq} = \sqrt{\frac{C_c}{C_a+C_{se}}}\, \left( \sqrt{\frac{R}{4} + 1} \pm \sqrt{\frac{R}{4}}\, \right)
\end{eqnarray}
and
\begin{eqnarray}\label{annihilation-R}
\Gamma_A^\chi = \frac{1}{2}\, \frac{C_c C_a}{C_a + C_{se}}\, \left( \sqrt{\frac{R}{4} + 1} \pm \sqrt{\frac{R}{4}}\, \right)^2\,,
\end{eqnarray}
where $R \equiv (C_s-C_e)^2/[C_c(C_a+C_{se})]$ is a dimensionless parameter defined by 5 coefficients in the DM evolution in Eq.~(\ref{N-evolution}), and the positive and negative signs are taken for $C_s > C_e$ and $C_s < C_e$, respectively~\cite{Chen:2014oaa}.
Using our numerical code, we can obtain the results consistent with those in Ref.~\cite{Chen:2014oaa}.

\begin{figure}
\begin{center}
\includegraphics[width=0.70\linewidth]{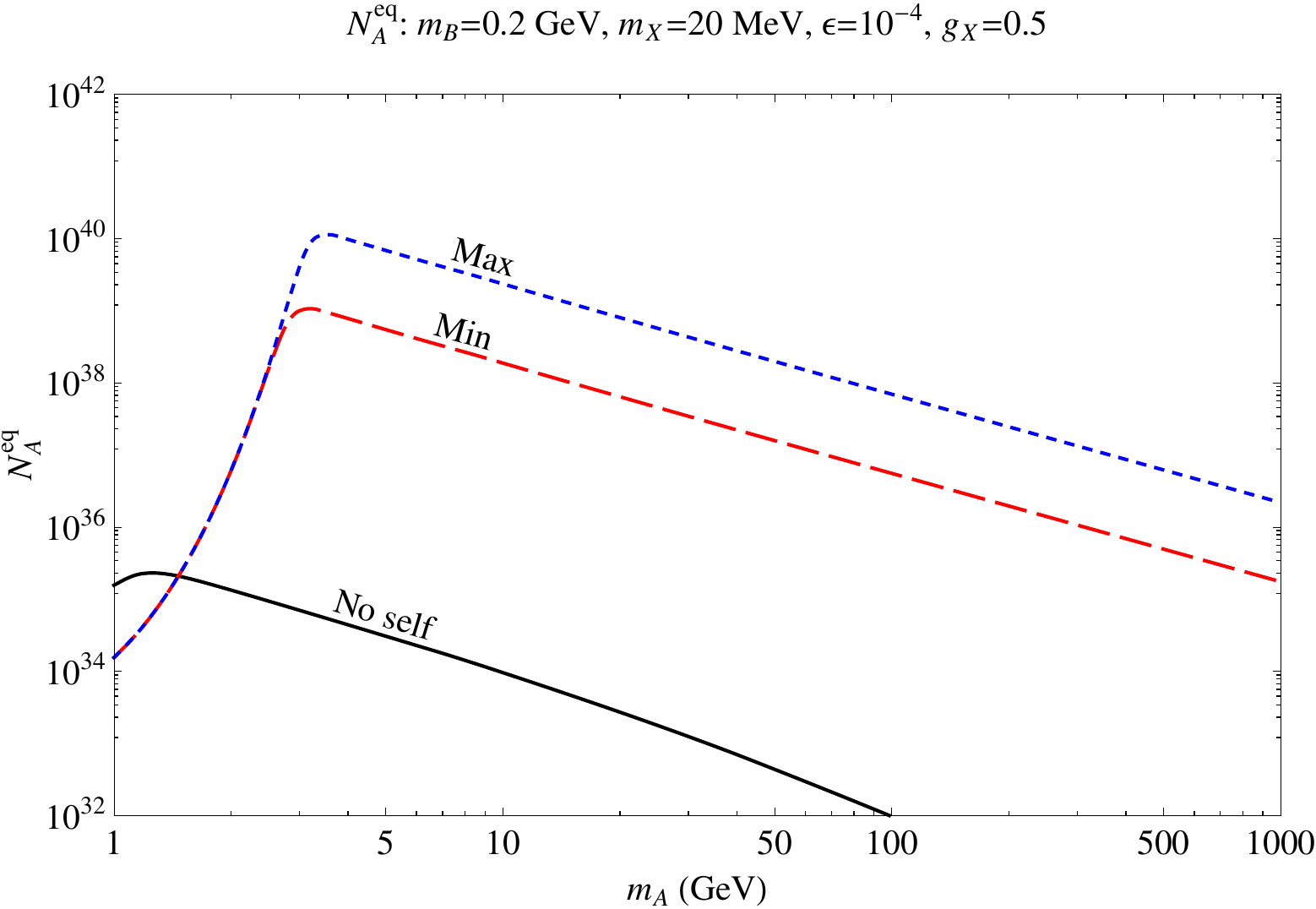}
\end{center}
\vspace*{-0.7cm}
\caption{Number of $\psi_A$ captured inside the Sun as a function of the $\psi_A$ mass $m_A$ for the benchmark parameters in Eq.~(\ref{benchmark}).
Each curve corresponds to No ($\sigma_{AA}/m_A = 0$), Min ($\sigma^{\rm min}_{AA}/m_A = 0.1\,{\rm cm}^2/{\rm g}$),
and Max ($\sigma^{\rm max}_{AA}/m_A =  1.25\,{\rm cm}^2/{\rm g}$) self-interaction, respectively.
}
\label{FigNeq}
\end{figure}

In Figure~\ref{FigNeq}, we present the number of heavy DM $\psi_A$ captured inside the Sun,
$N_A^{\rm eq}$, for the benchmark model parameters as in Eq.~(\ref{benchmark}).
Min and Max curves respectively correspond to minimum and maximum values of the self-interaction of $\psi_A$ in the preferred range, $0.1\,{\rm cm}^2/{\rm g} < \sigma_{AA}/m_A < 1.25\,{\rm cm}^2/{\rm g}$~\cite{Spergel:1999mh, Rocha:2012jg, Randall:2007ph, Zavala:2012us}.
In the case of no self-interaction, the amount of accumulated $\psi_A$ is quite small since the $\psi_A$-nucleon scattering cross section is suppressed as explained in Section~\ref{XSection}.
However, the self-interaction of $\psi_A$, $\sigma_{AA}$ can significantly enhance $N_A^{\rm eq}$.
%
%

\subsection{Flux of boosted dark matter}

The flux of boosted DM $\psi_B$ from the Sun through the annihilation $\psi_A\overline{\psi}_A \to \psi_B\overline{\psi}_B$ can be expressed as
\begin{eqnarray}
\frac{d\Phi_B^{\rm Sun}}{dE_B} = \frac{\Gamma_A^{\psi_A}}{4\pi R_{\rm Sun}^2}\, \frac{dN_B}{dE_B}\,,
\label{eq:flux}
\end{eqnarray}
where $R_{\rm Sun}$ is the distance between the Sun and the Earth, $\Gamma_A^{\psi_A}$ is the annihilation rate of heavy DM $\psi_A$ in the Sun, and $dN_B/dE_B$ is the differential energy spectrum of boosted DM $\psi_B$ at the source.
The differential spectrum is simply given by
\begin{eqnarray}
\frac{dN_B}{dE_B} = 2 \delta(E_B - m_A) \, ,
\end{eqnarray}
since the annihilation of heavy DM $\psi_A$, $\psi_A\overline{\psi}_A \to \psi_B\overline{\psi}_B$ produces two mono-energetic boosted $\psi_B$'s.
The annihilation rate of $\psi_A$ in the Sun, $\Gamma_A^{\psi_A}$, is obtained from Eq.~(\ref{annihilation}) (or Eq.~(\ref{annihilation-R})) with Eqs.~(\ref{N-final}) and (\ref{Equi-time}) for $t = t_\odot$.
Note that there is no need to consider the line-of-sight integration in Eq. (\ref{eq:flux}),
since the annihilation $\psi_A\overline{\psi}_A \to \psi_B\overline{\psi}_B$ in the Sun provides a point-like source of the boosted DM $\psi_B$.
This is different from the case with the boosted DM flux from the GC as in Ref.~\cite{Agashe:2014yua},
where one needs to compute a halo-dependent integral over the line-of-sight.

\subsection{Energy loss in the Sun}\label{EnergyLoss}

The boosted DM particles $\psi_B$ produced from the annihilation $\psi_A\overline{\psi}_A \to \psi_B\overline{\psi}_B$ in the Sun may lose their kinetic energy as they pass through the Sun from their production points due to the relatively large scattering cross section with electrons, $\sigma_{Be^- \to Be^-}$ and the large radius of the Sun, $R_\odot \simeq 6.96\times 10^{10}\, {\rm cm}$.
The energy loss of the particles during propagation through matter is well discussed in Ref.~\cite{Agashe:2014kda}.
The boosted DM particle propagating through matter loses its energy dominantly through ionization of atoms,
which is very similar to the energy loss of a heavy charged SM particle \cite{Agashe:2014yua}.
For $\beta\gamma=p/Mc$ around the range of $\mathcal{O}(10-100)$, the mean rate of energy loss of a muon is $\sim1$ GeV/m inside the Earth and $\sim0.6$ GeV/m inside the Sun.
The boosted DM $\psi_B$ scatters off SM particles via a $t-$channel $X$ boson exchange while the muon does via a $t-$channel photon exchange.
Analogous to Ref.~\cite{Agashe:2014yua}, we can easily approximate the required travel length for the $\psi_B$ to lose 1 GeV of energy by comparing the couplings and propagator of the $\psi_B-e$ scattering and those of the $\mu-e$ scattering:
\begin{eqnarray}
L_{\psi_B}^{\rm Sun} \approx L_\mu^{\rm Sun}\, \left[ \frac{\epsilon^2 g_X^2}{e^2}\,
\left( \frac{t}{t - m_X^2} \right)^2 \right]^{-1}\,,
\end{eqnarray}
where $t=2m_e(m_e-E_e)$ and $L_\mu^{\rm Sun} \simeq (100/0.6)$ cm.

%
\begin{figure}
\begin{center}
\includegraphics[width=0.70\linewidth]{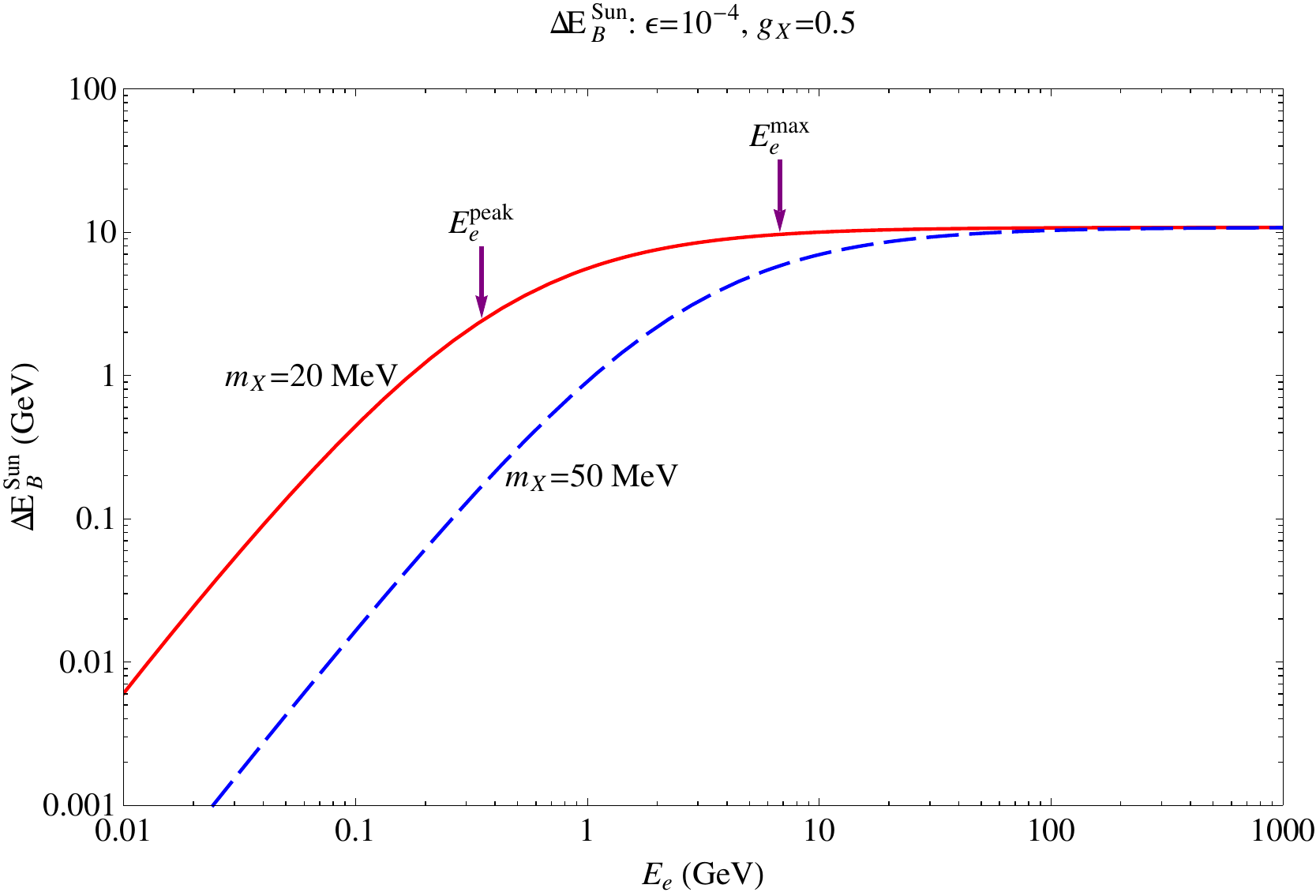}
\end{center}
\vspace*{-0.7cm}
\caption{Required energy for $\psi_B$ to escape from the Sun as a function of the scattered electron energy $E_e$ for $m_X = 20$ MeV and 50 MeV.
The kinetic mixing and hidden coupling are fixed as $\epsilon=10^{-4}$ and $g_X=0.5$.
The purple arrows indicate the $E_e^{\rm peak}$ and $E_e^{\rm max}$ values for the benchmark scenario in Eq.~(\ref{benchmark}).
}
\label{FigDEB}
\end{figure}
%

For the benchmark scenario in Eq.~(\ref{benchmark}), we estimate $L_{\psi_B}^{\rm Sun} \approx 3 \times 10^{10}\, (0.5/g_X)^2(10^{-4}/\epsilon)^2$ cm which is about a factor of 2 smaller than $R_\odot \simeq 6.96\times 10^{10}\, {\rm cm}$.
To escape from the Sun, the boosted DM $\psi_B$ of the benchmark scenario will lose $\sim 2$ GeV of energy on average which corresponds to $\sim 10 \%$ of the initial energy of $\psi_B$, $E_B^i\simeq m_A = $ 20 GeV.
For the above estimation, we use $E_e=E_e^{\rm peak}$, the electron energy corresponding to the peak of the recoil electron spectrum which is a reasonable choice since the $\psi_B-e$ scattering mostly occurs around
the peak energy in the electron recoil spectrum.
For comparison, we obtain $L_{\psi_B}^{\rm Sun} \approx 7 \times 10^{9}\, (0.5/g_X)^2(10^{-4}/\epsilon)^2$ cm for the most extreme (conservative) case $E_e=E_e^{\rm max}$.
In the following section, we will numerically compute the required energy of $\psi_B$, $\Delta E_B^{\rm Sun}$, to escape from the production point to the surface of the Sun assuming that the travel distance of $\psi_B$ inside the Sun is equal to the radius of the Sun, $R_\odot$.
In Figure~\ref{FigDEB}, we show $\Delta E_B^{\rm Sun}$ as a function of the scattered electron energy $E_e$ for two representative cases, $m_X = 20$ MeV (solid red) and 50 MeV (blue dashed).
The kinetic mixing and hidden coupling are taken from the benchmark parameter set in Eq.~(\ref{benchmark}): $\epsilon=10^{-4}$ and $g_X=0.5$.
As shown in the Figure, the $\Delta E_B^{\rm Sun}$ becomes smaller for larger $m_X$ and smaller $E_e$,
and converges to $\sim 10\, {\rm GeV}\, (g_X/0.5)^2 (\epsilon/10^{-4})^2 $ when $|t| \gg m_X^2$, i.e., $E_e \gg m_X^2/m_e$.
The ratio between the energy loss and the initial energy of $\psi_B$, $\Delta E_B^{\rm Sun}/E_B^i$ can be larger than $0.1$, even $\mathcal{O}(1)$, for a low $\psi_A$ mass ($m_A \lesssim 10$ GeV), i.e. small $E_e$.
Consequently, in our analysis, we use $E_B^f = E_B^i - \Delta E_B^{\rm Sun} = m_A-\Delta E_B^{\rm Sun}$ as the energy of $\psi_B$ in a detector
and the parameter region for $E_B^i < \Delta E_B^{\rm Sun}$ is not scanned.

\section{Detection of boosted dark matter}
\label{sec:detection}

In this section, we discuss detection prospects of boosted DM particles in neutrino detection experiments.
Because of the relatively small flux and weak interaction of boosted DM,
large volume experiments such as Super-K, Hyper-K, and PINGU are preferred.
Large volume neutrino experiments have been developed to detect energetic charged particles scattered off from neutrino-matter collisions.
Such energetic charged particles (electrons in this analysis) may be generated through scattering with the boosted DM $\psi_B$ from the Sun.
We can easily reduce the number of background events with a better angular resolution on the Cherenkov-emitted electron direction since we are interested in the flux of boosted DM from a point-like source, the Sun.
Thus, an angular resolution of each experiment is very crucial in this analysis.
Moreover, the electron energy $E_e$ shows a peak in its recoil spectrum at relatively low values due to the $t-$channel $X$ boson~\cite{Agashe:2014yua}.
Thus, Super-K and Hyper-K are very well fitted experiments to detect the boosted DM flux from the Sun due to their good angular resolution $\theta_{\rm res} \simeq 3^\circ$ and low energy threshold $E_e^{\rm th} \simeq 0.01$ GeV.
Although PINGU has a higher energy threshold $E_e^{\rm th} \simeq 1$ GeV and worse angular resolution $\theta_{\rm res} \simeq 23^\circ$, it will be able to have some sensitivity as shown in the following subsections.
We will not discuss IceCube in spite of its very large volume ($\sim10^3$ Mton) due to its high energy threshold, $E_e^{\rm th} > 100$ GeV.
See Table 1 of Ref.~\cite{Agashe:2014yua} for a summary of relevant neutrino experiments.
A brief discussion on the detection prospects from the Earth is found in Section~\ref{Earth}.

\subsection{Signals}

As discussed in Ref.~\cite{Agashe:2014yua}, the signal of boosted DM $\psi_B$ can be detected mainly through its elastic scattering off electrons, $\psi_B e^- \to \psi_B e^-$.
Unlike the thermal relic $\psi_A$ around the GC, the $\psi_A$ trapped in the Sun becomes a point-like source of boosted DM $\psi_B$,
and thus we need no angular-cut, $\theta_C$.
Finally, the number of electron signal events is given by
\begin{eqnarray}
N_{\rm sig} &=& \Delta T\, N_{\rm target}\, \Phi_B^{\rm Sun}\, \sigma_{Be^- \to Be^-}\\
&=& \Delta T\, \frac{10\, \rho_{\rm target}\,
V_{\rm exp}}{m_{{\rm H}_2{\rm O}}}\,
\frac{2 \Gamma_A^{\psi_A}}{4\pi R_{\rm Sun}^2}\,
\int_{E_e^{\rm min}}^{E_e^{\rm max}} dE_e\,
\frac{d \sigma_{Be^- \to Be^-}}{d E_e}\,,
\end{eqnarray}
where $\Delta T$ is the exposure time of the measurement, $N_{\rm target}$ is the total number of target electrons, $\Phi_B^{\rm Sun}$ is the boosted DM flux from the Sun, $\sigma_{Be^- \to Be^-}$ is the $\psi_B-e$ scattering cross section, and the factor of 10 in the second line is the number of electrons per water molecule.
In order to avoid backgrounds from solar neutrinos~\cite{SolarModelFile} and muons, we use a minimum energy-cut $E_e^{\rm min}=0.1$ GeV instead of $E_e^{\rm th} \simeq 0.01$ GeV for Super-K and Hyper-K following Ref.~\cite{Agashe:2014yua} while $E_e^{\rm min} = E_e^{\rm th} \simeq 1$ GeV is enough for PINGU.

%
\begin{figure}
\begin{center}
\includegraphics[width=0.322\linewidth]{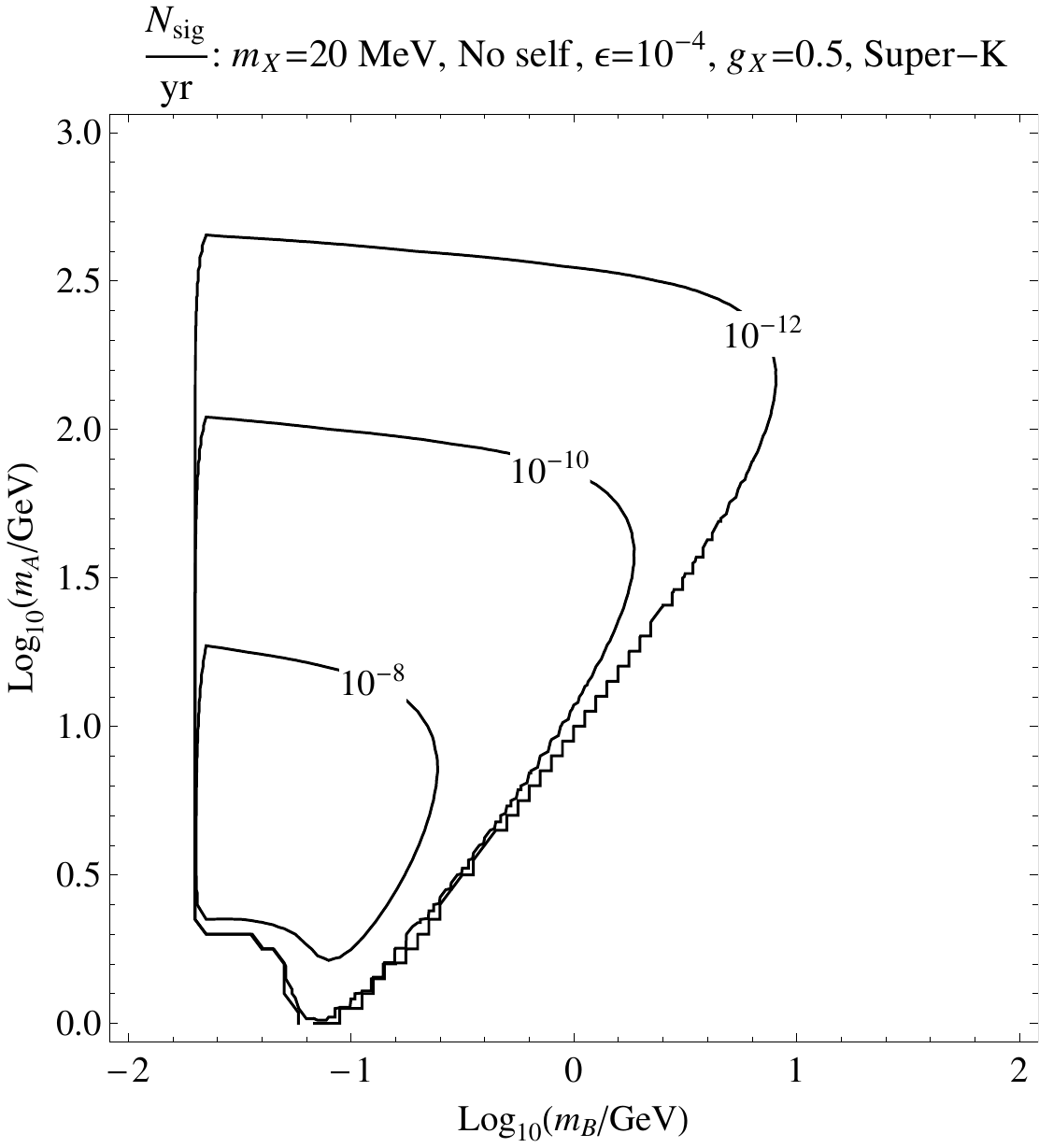}
\hspace*{0.03cm}
\includegraphics[width=0.322\linewidth]{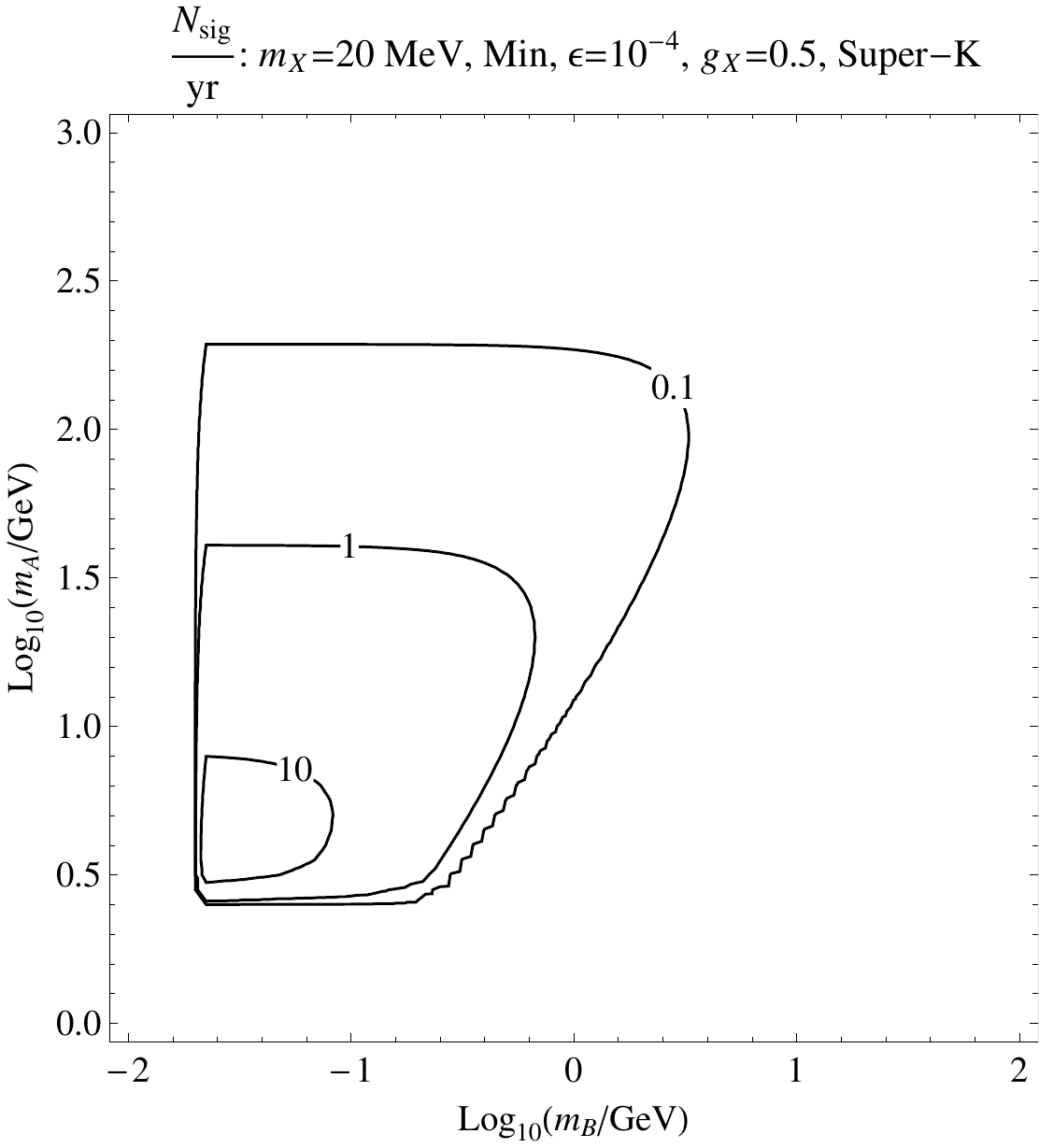}
\hspace*{0.03cm}
\includegraphics[width=0.322\linewidth]{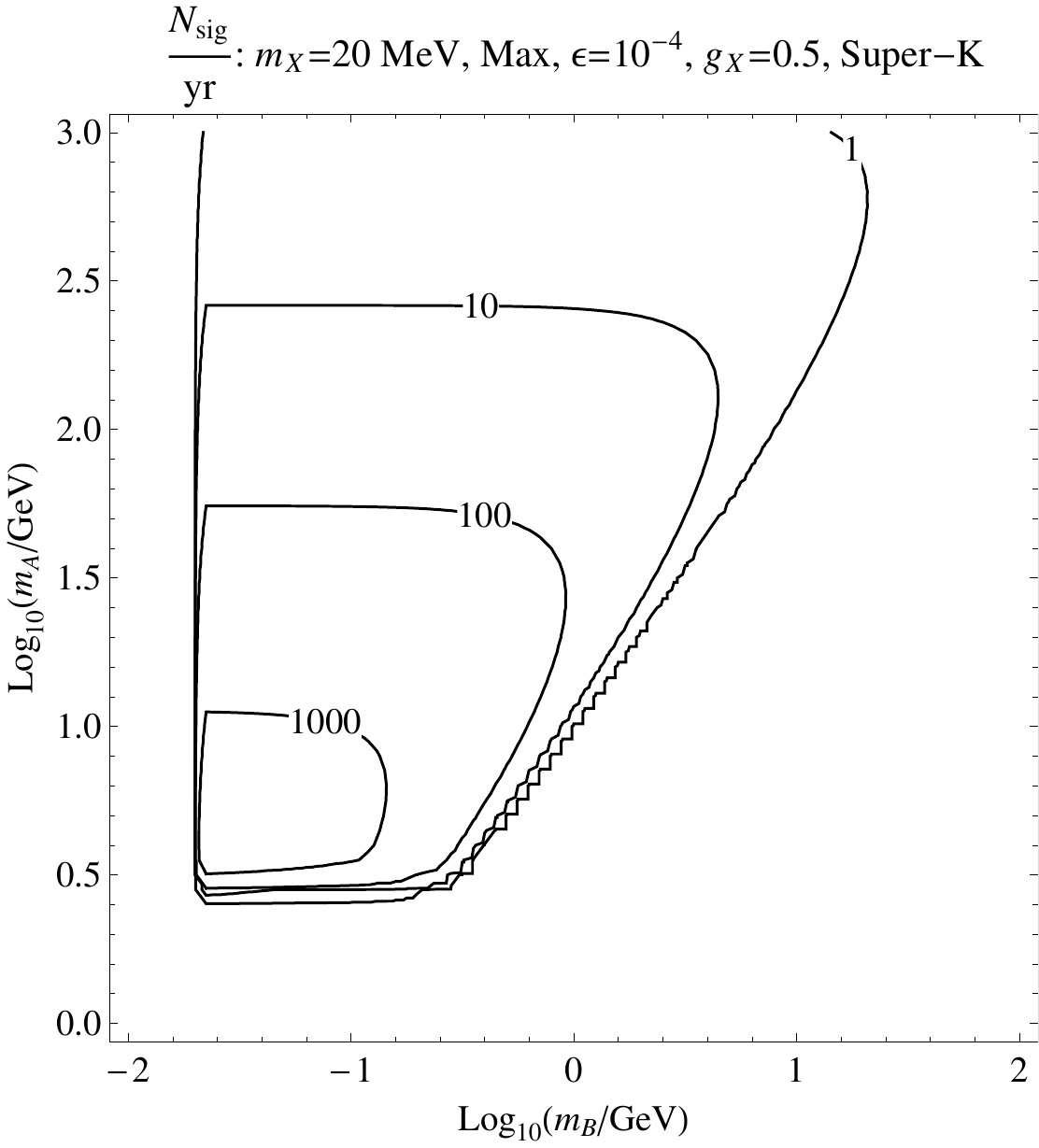}
\vspace*{-0.4cm}

\includegraphics[width=0.322\linewidth]{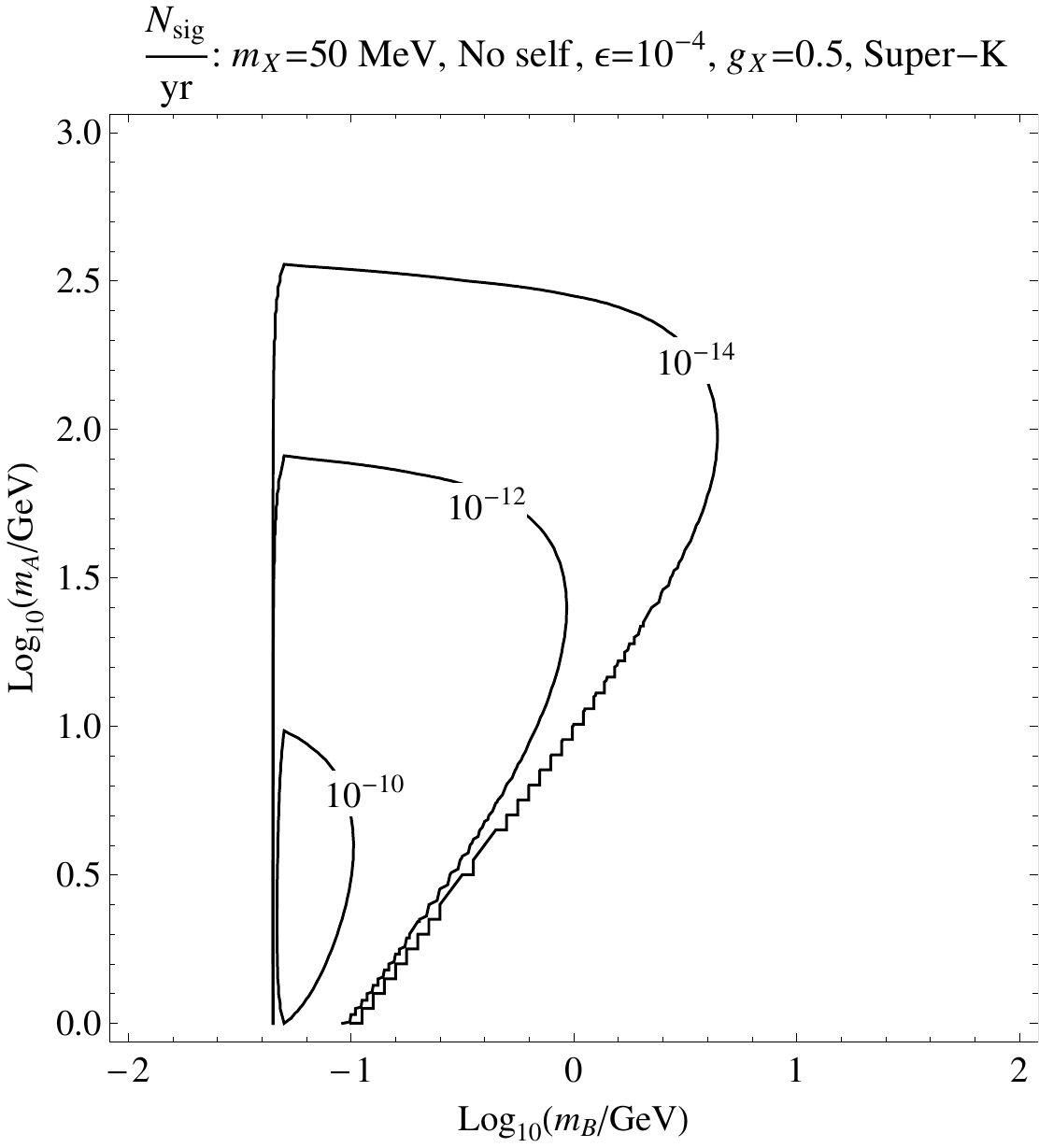}
\hspace*{0.03cm}
\includegraphics[width=0.322\linewidth]{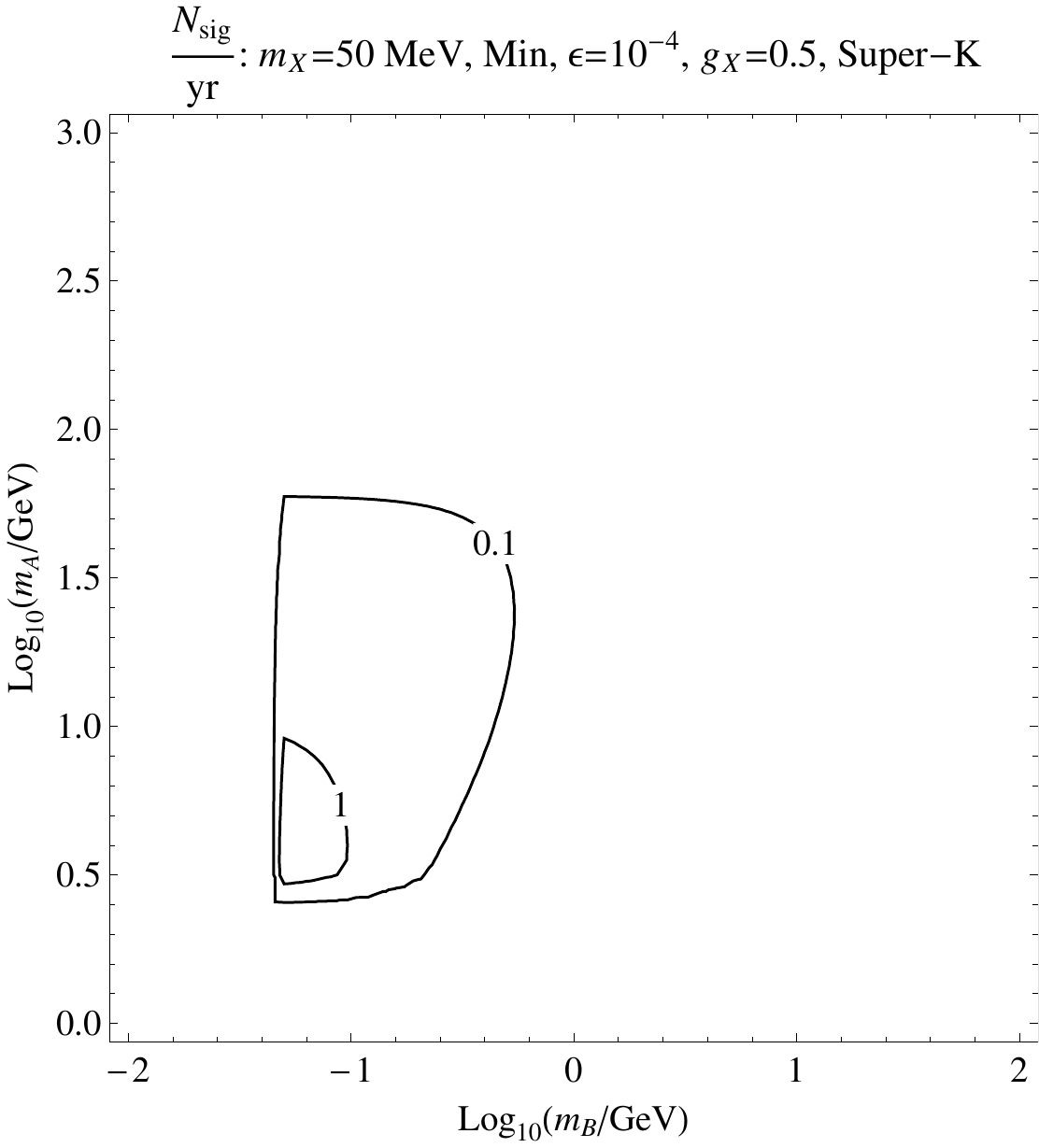}
\hspace*{0.03cm}
\includegraphics[width=0.322\linewidth]{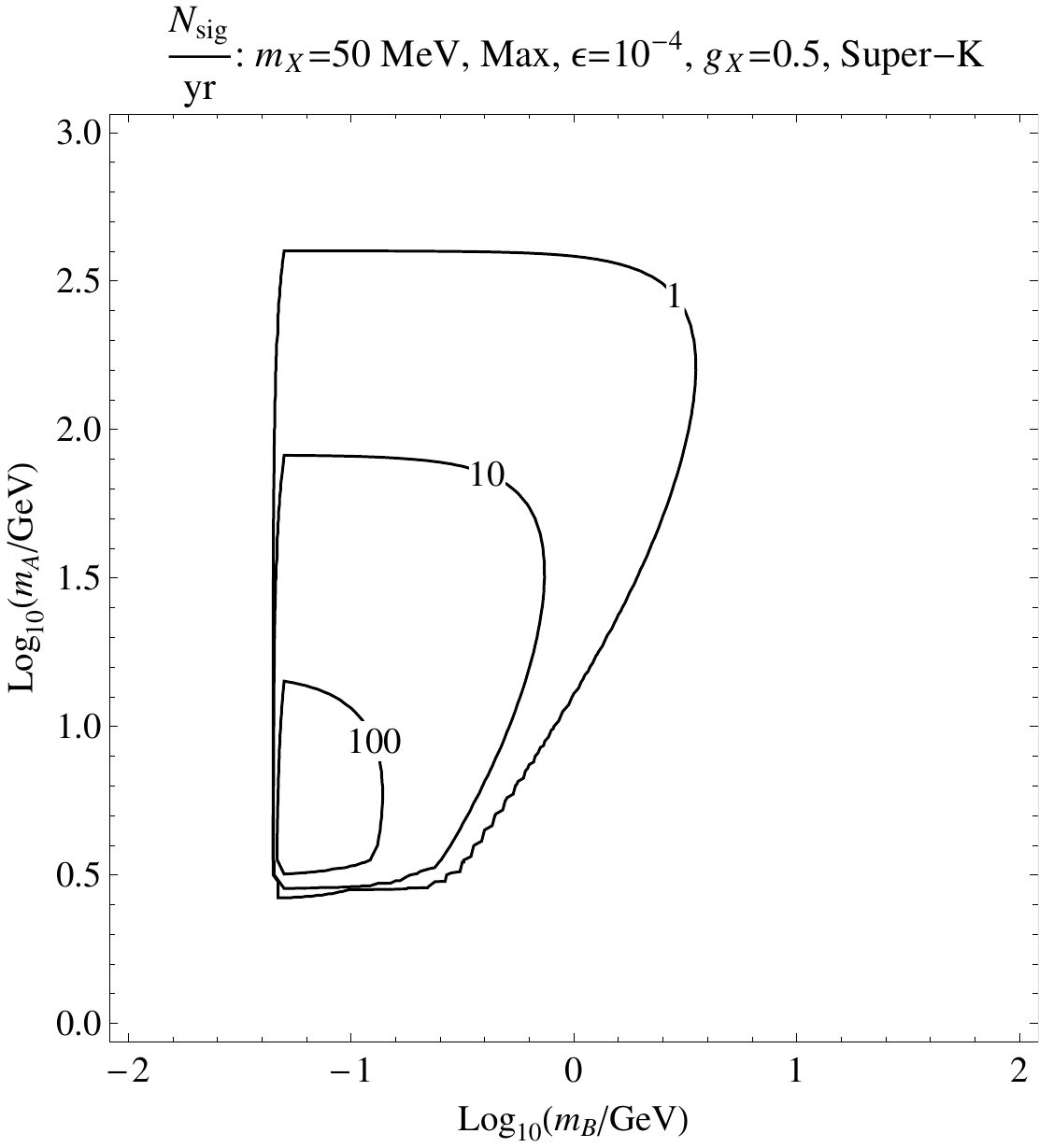}
\end{center}
\vspace*{-0.7cm}
\caption{
Number of signal events per year in Super-K in the ($m_B$, $m_A$) plane
for  $m_X=20$ MeV (top) and 50 MeV (bottom) and
No, Min, and Max self-interaction (left to right), respectively.
The kinetic mixing and hidden coupling are fixed as the benchmark parameters in Eq.~(\ref{benchmark}), $\epsilon=10^{-4}$ and $g_X=0.5$.
}
\label{Fig1}
\end{figure}
%

%
\begin{figure}
\begin{center}
\includegraphics[width=0.40\linewidth]{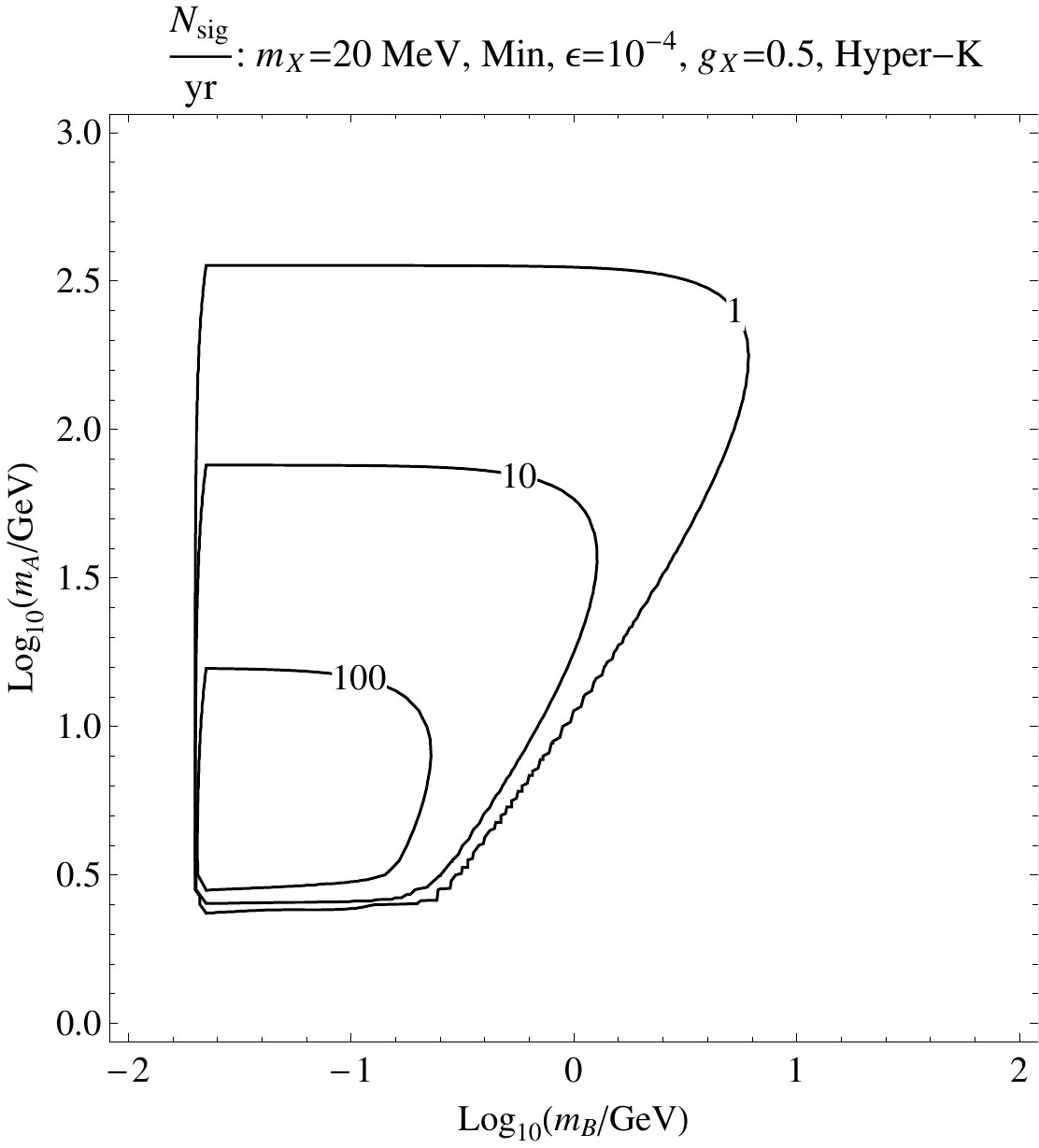}
\hspace*{0.7cm}
\includegraphics[width=0.40\linewidth]{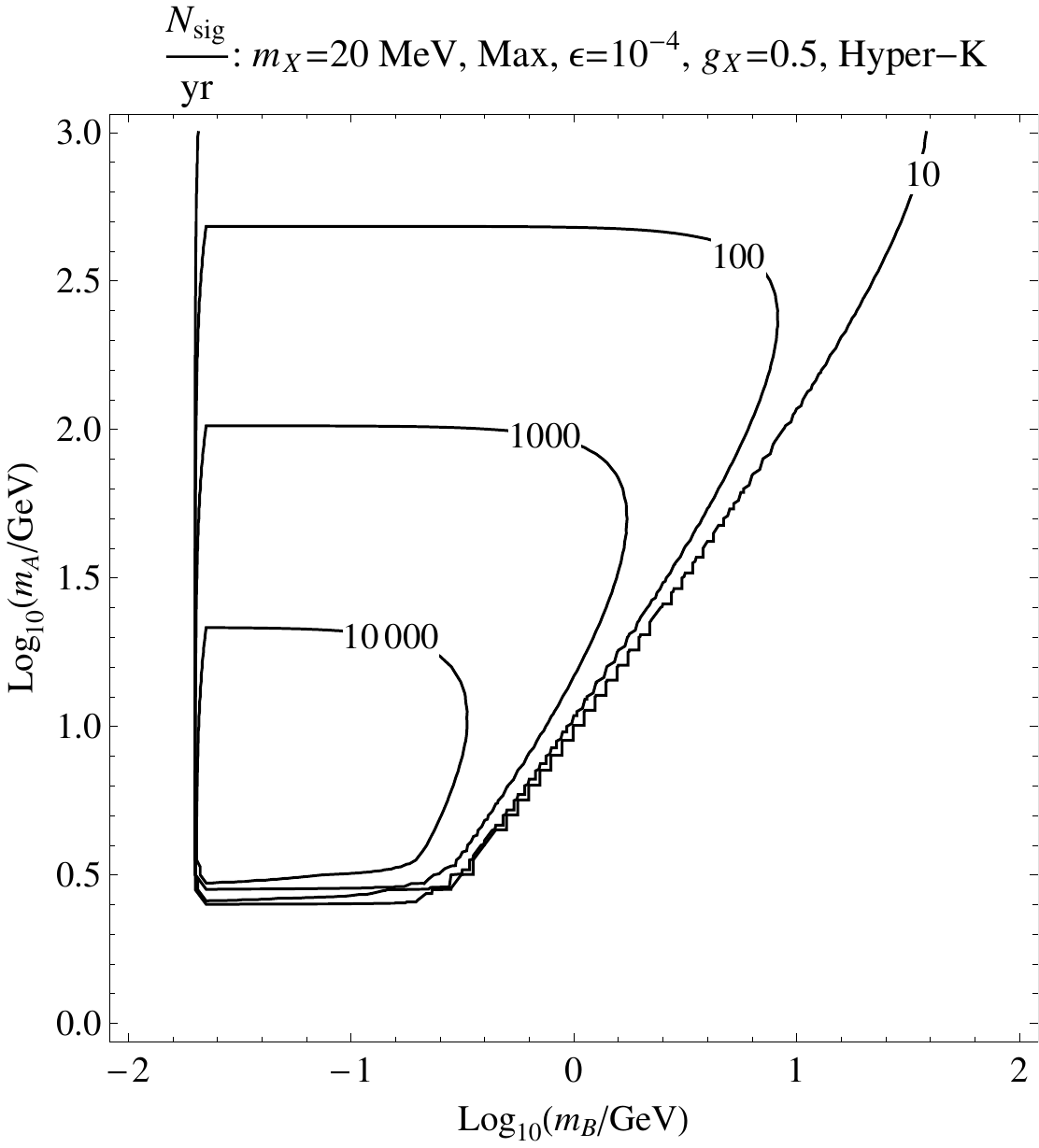}
\vspace*{0.4cm}

\includegraphics[width=0.40\linewidth]{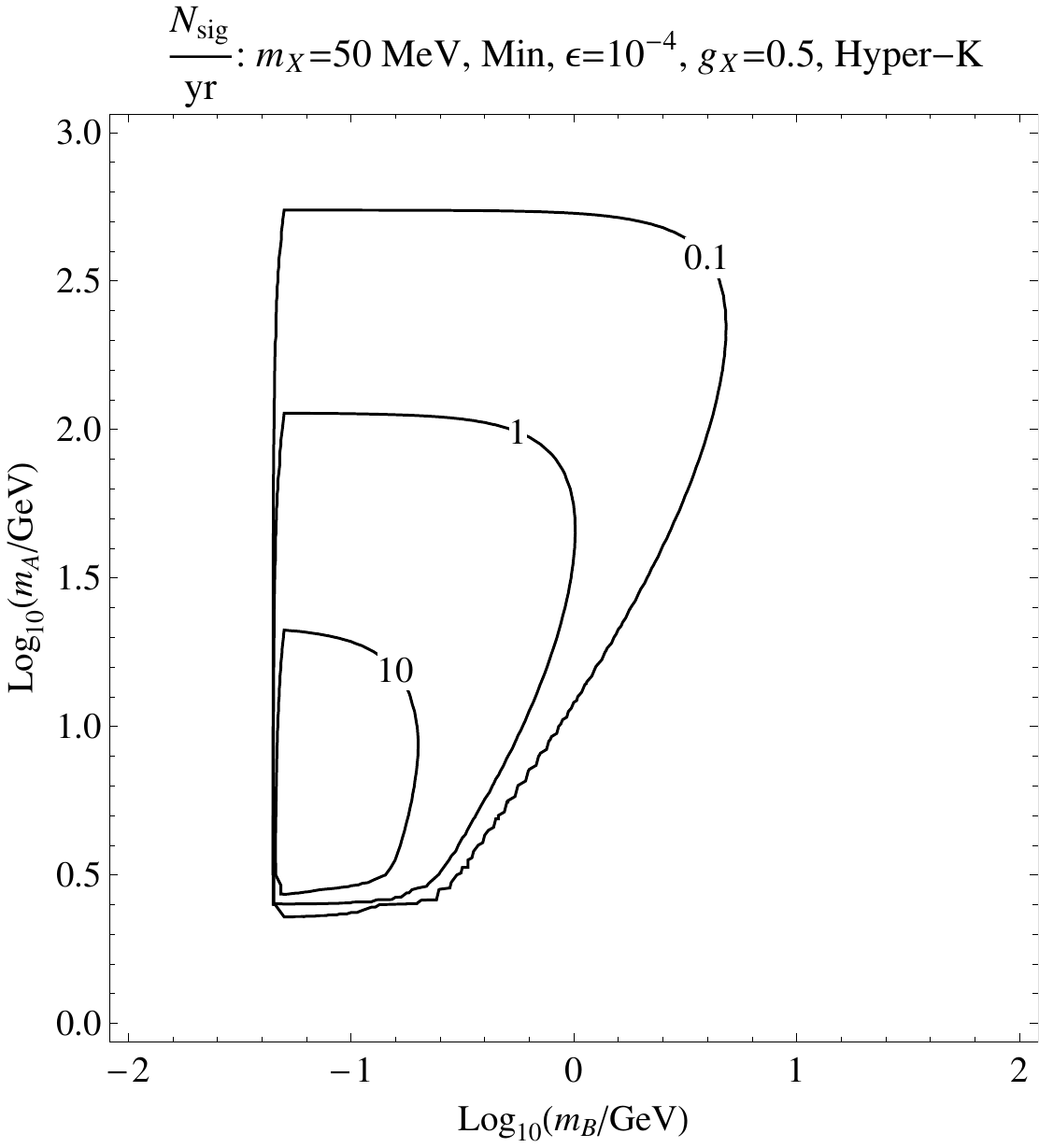}
\hspace*{0.7cm}
\includegraphics[width=0.40\linewidth]{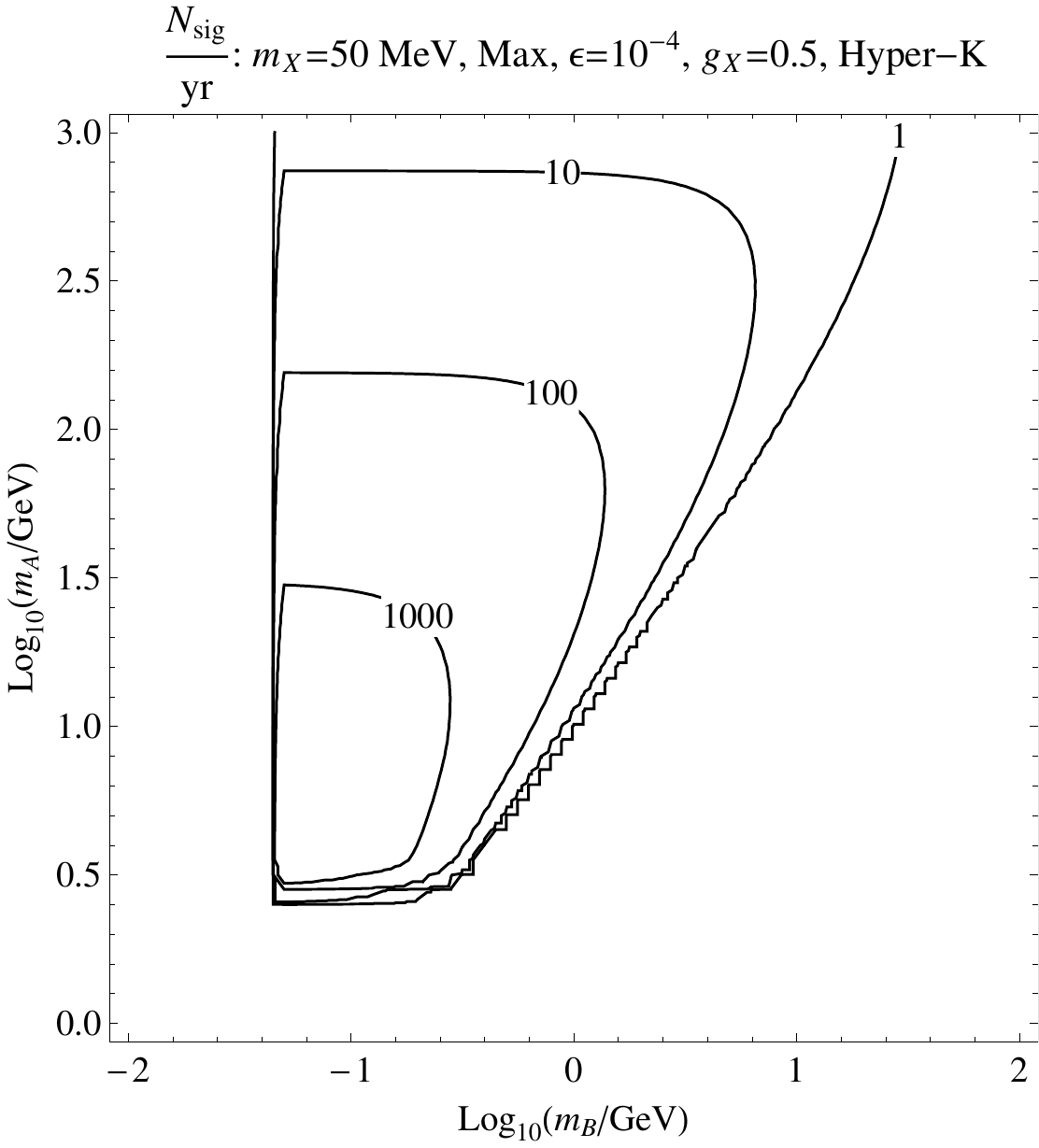}
\end{center}
\vspace*{-0.7cm}
\caption{Same as in Figure \ref{Fig1} for Min (left) and Max (right) self-interaction, but for Hyper-K.
}
\label{Fig2}
\end{figure}
%

%
\begin{figure}
\begin{center}
\includegraphics[width=0.40\linewidth]{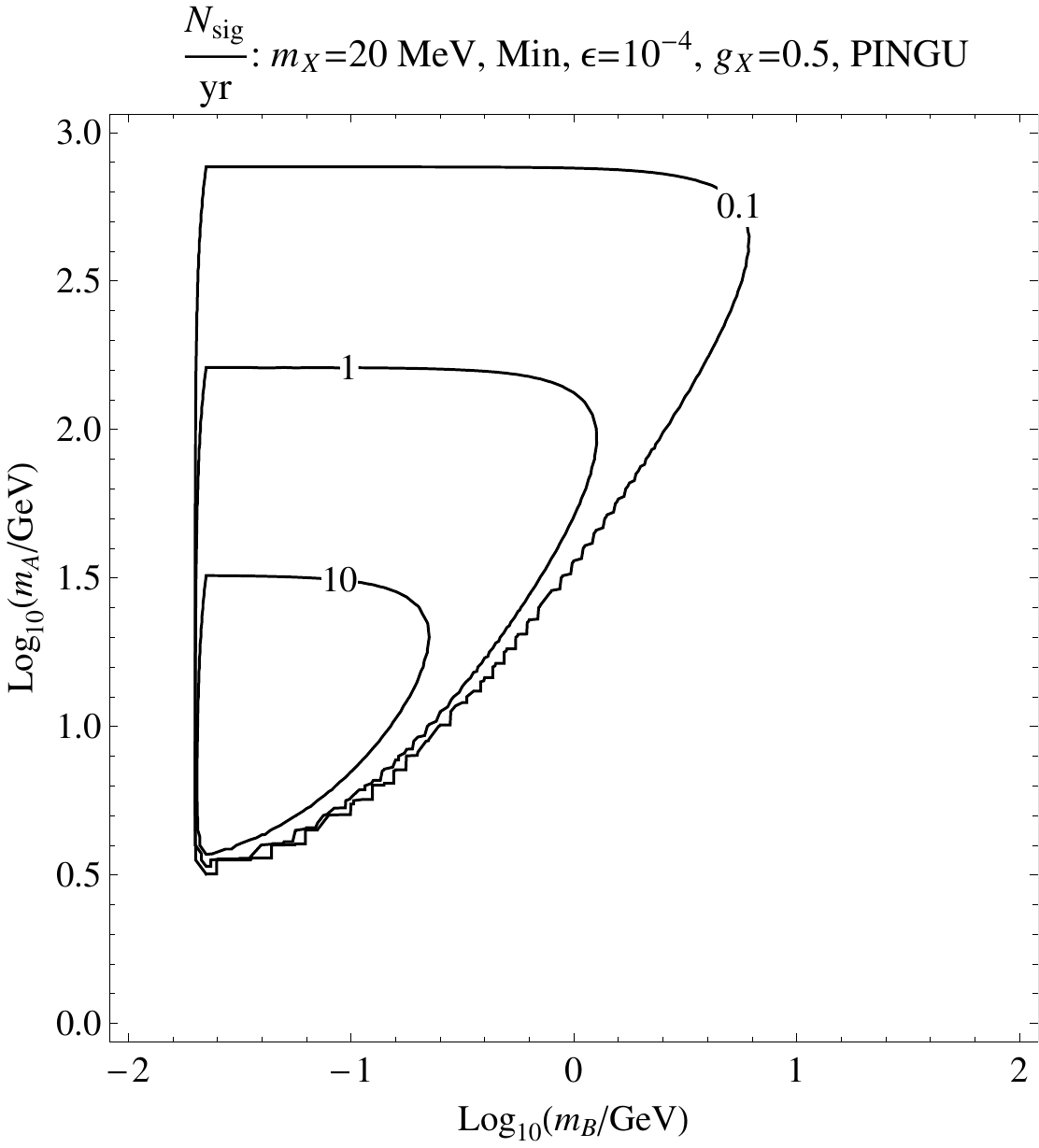}
\hspace*{0.7cm}
\includegraphics[width=0.40\linewidth]{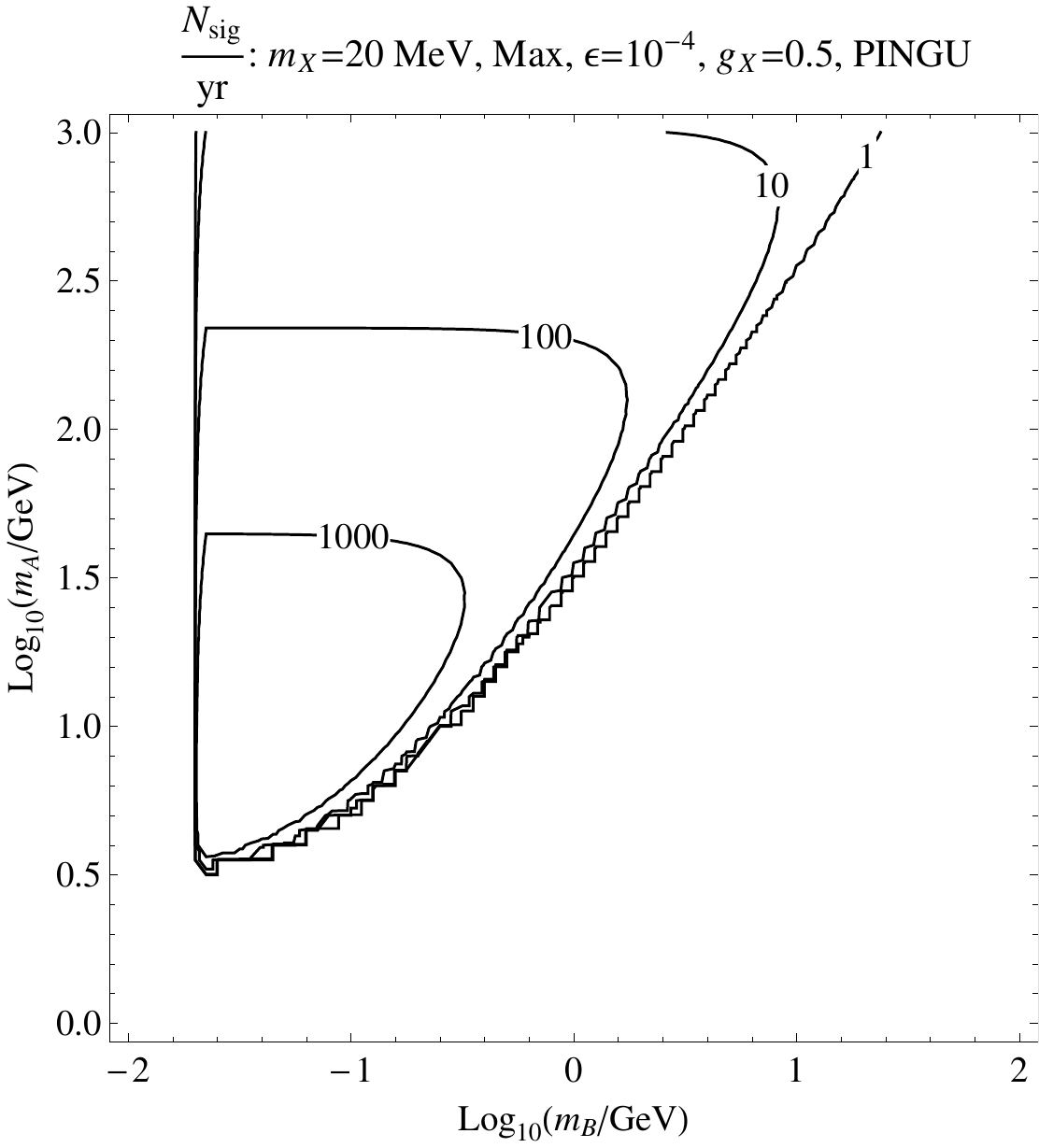}
\vspace*{0.4cm}

\includegraphics[width=0.40\linewidth]{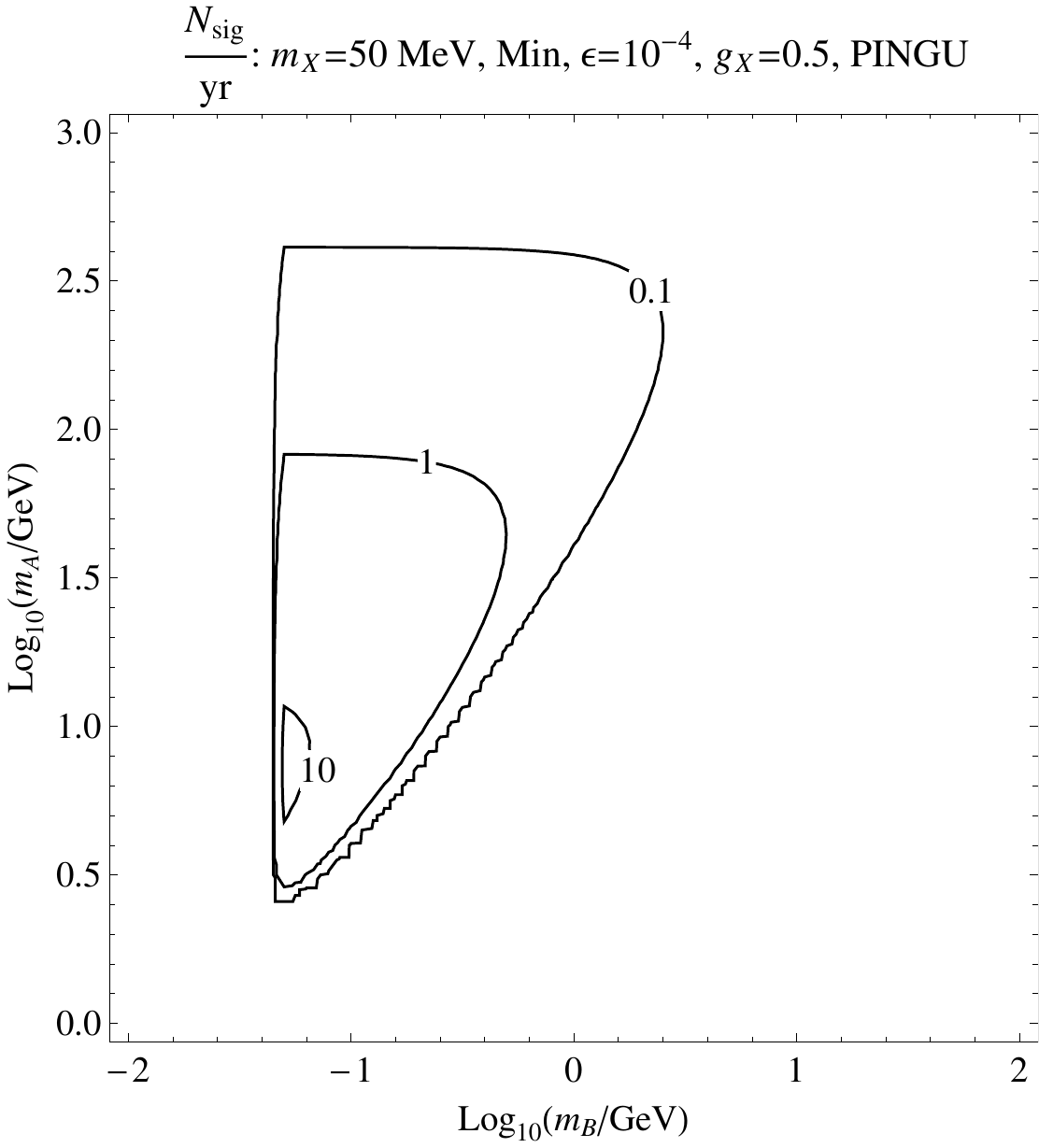}
\hspace*{0.7cm}
\includegraphics[width=0.40\linewidth]{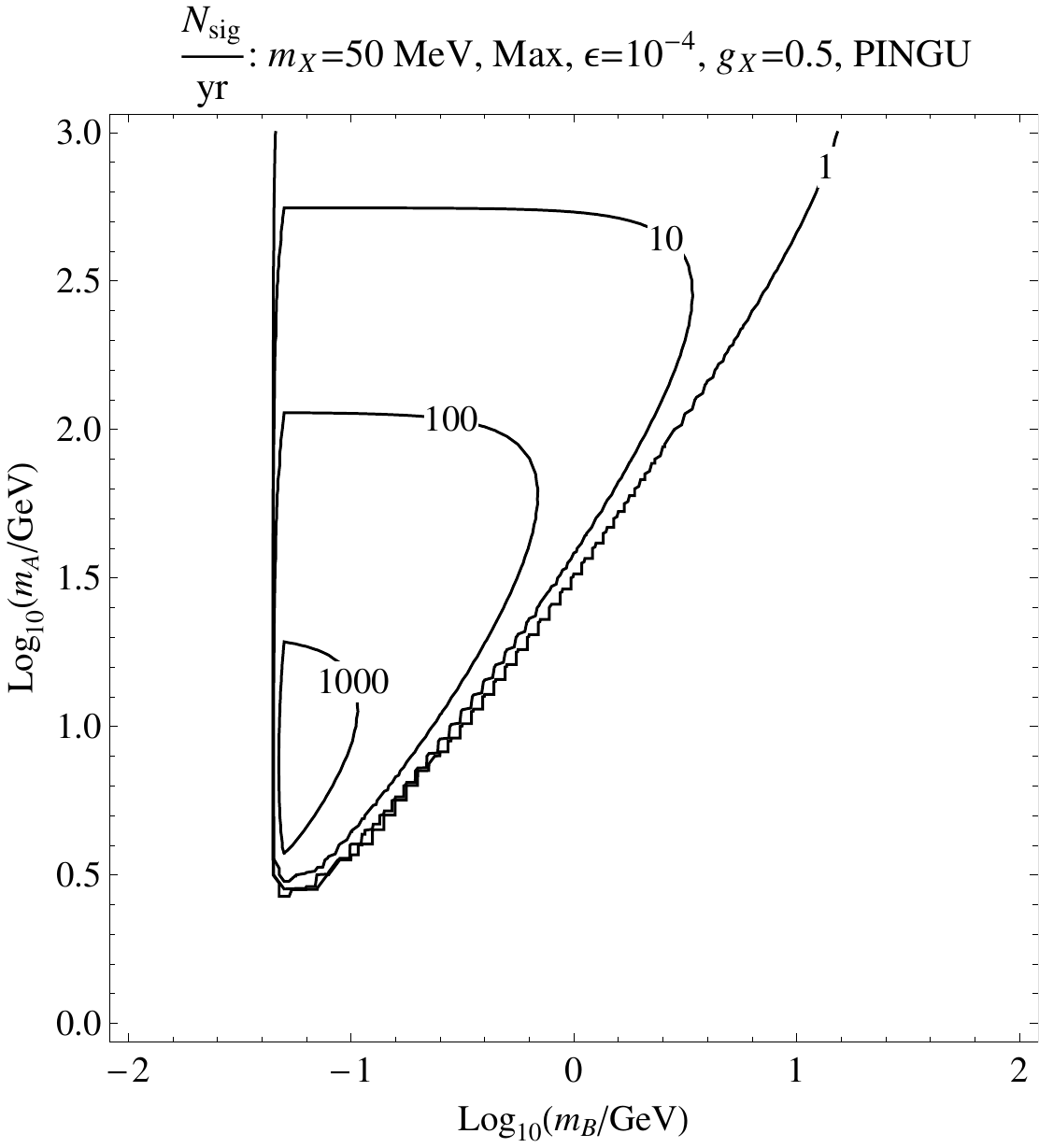}
\end{center}
\vspace*{-0.7cm}
\caption{Same as in Figure \ref{Fig1} for Min (left) and Max (right) self-interaction, but for PINGU.
}
\label{Fig3}
\end{figure}
%

In Figures~\ref{Fig1}, \ref{Fig2}, and \ref{Fig3}, we show the number of signal events per year in the ($m_B$, $m_A$) plane for three experiments: Super-K, Hyper-K, and PINGU, respectively.
For each Figure, we use benchmark values: $m_X=20$ MeV (top) and 50 MeV (bottom) and No (only for Super-K), Min, and Max self-interaction (left to right).
The kinetic mixing and hidden coupling are fixed as the benchmark values in Eq.~(\ref{benchmark}), $\epsilon=10^{-4}$ and $g_X=0.5$.
Naturally we can detect more signal events in an experiment with a larger volume and a lower $E_e^{\rm th}$, and also for stronger interactions of DM particles.
In the Figures, for the signal number contours the left-edge is set by the condition $m_B > m_X$, the top-edge is by the DM number density $\propto 1/m_{\rm DM}$, the right-diagonal-edge is by $E_e^{\rm max} > E_e^{\rm min}$ which is approximated as $m_A > 10\, (30)\times m_B$ for Super-K/Hyper-K (PINGU),
and the bottom-edge is by the rapid drop in the accumulated number of DM particles inside the Sun for $m_{\rm DM} \lesssim 3$ GeV (see Figure~\ref{FigNeq}) due to the very active evaporation effects.

%
\begin{figure}
\begin{center}
\includegraphics[width=0.40\linewidth]{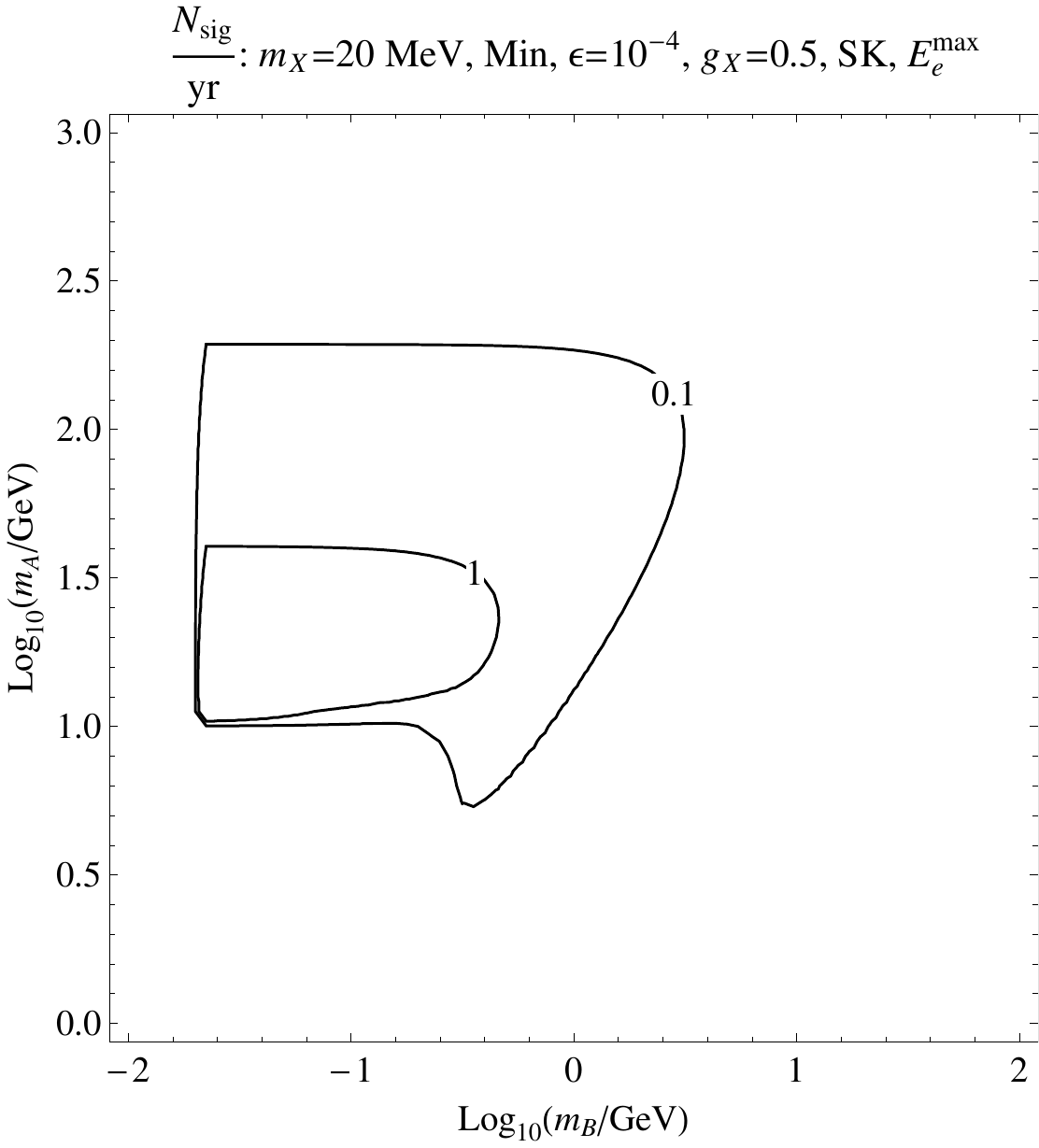}
\hspace*{0.7cm}
\includegraphics[width=0.40\linewidth]{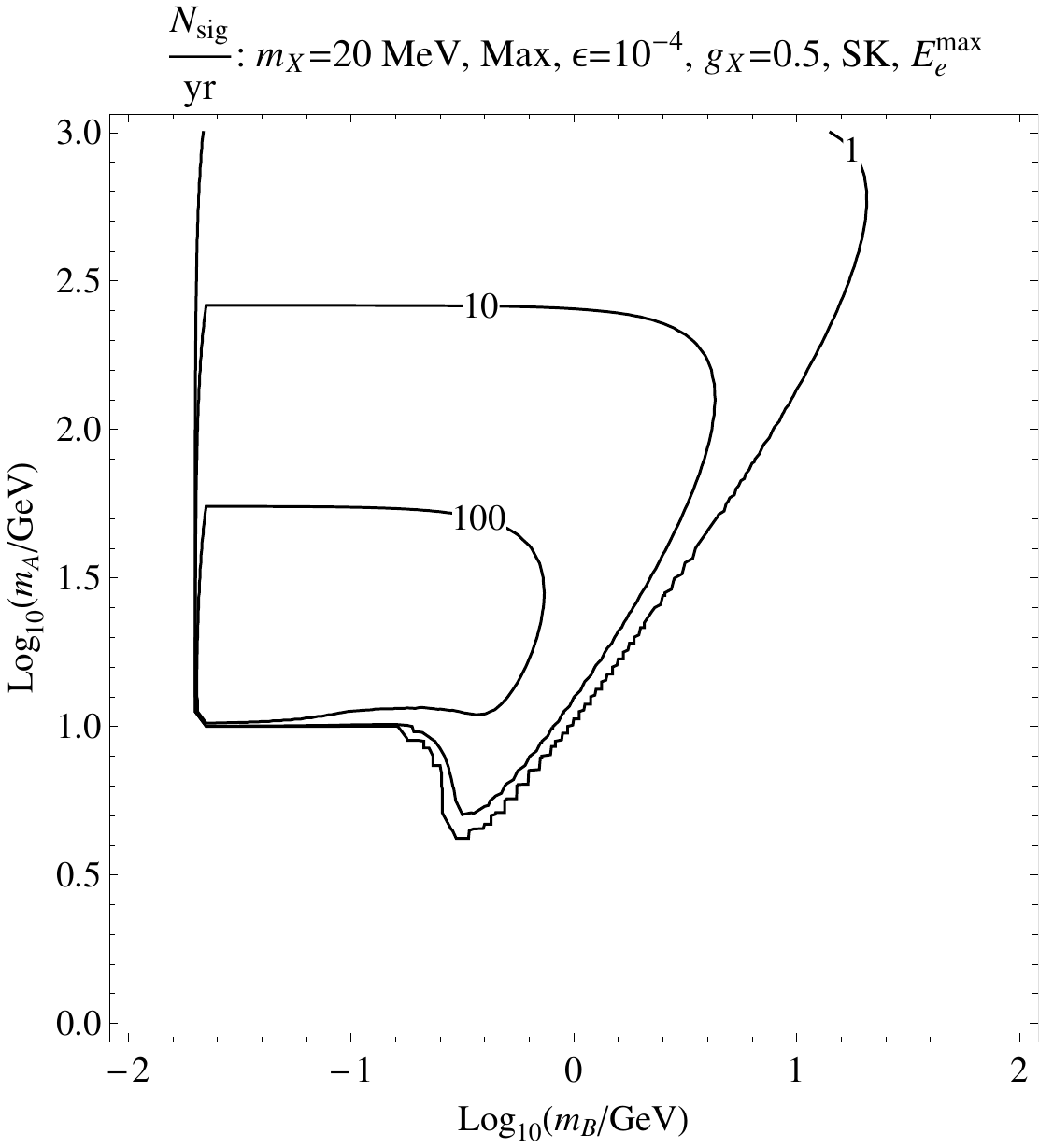}
\vspace*{0.4cm}

\includegraphics[width=0.40\linewidth]{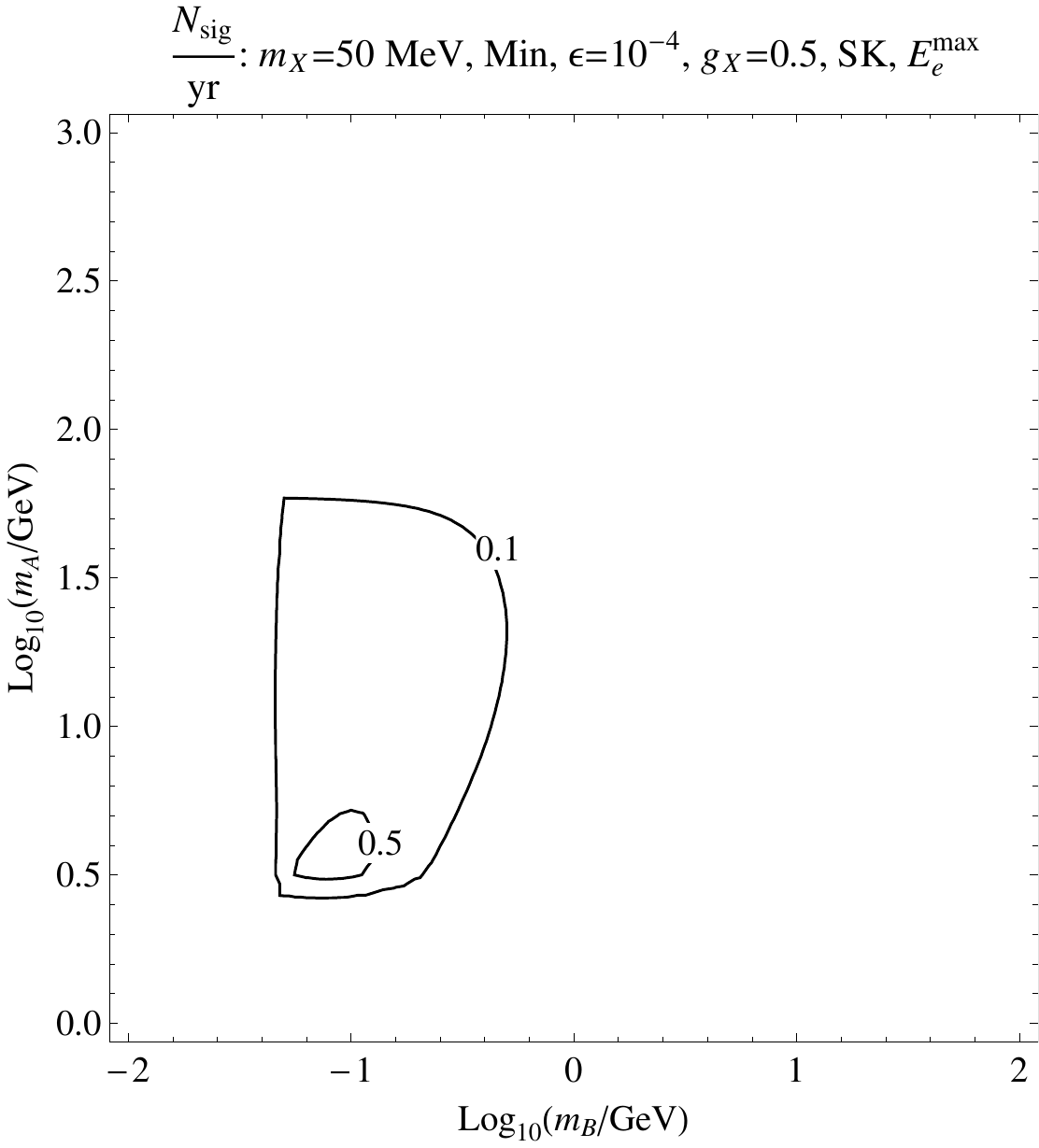}
\hspace*{0.7cm}
\includegraphics[width=0.40\linewidth]{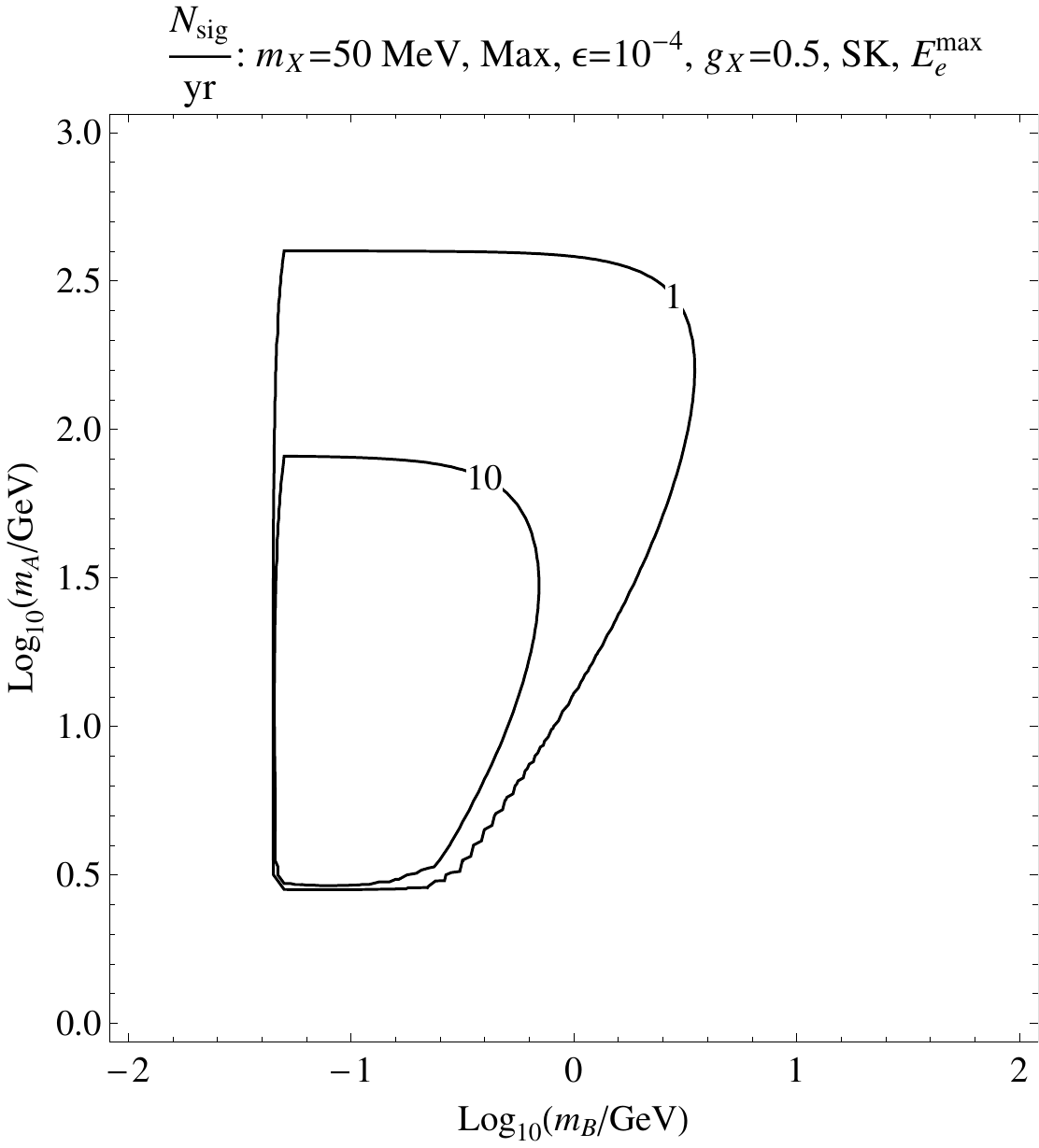}
\end{center}
\vspace*{-0.7cm}
\caption{Same as in Figure \ref{Fig1} for Min (left) and Max (right) self-interaction,
but $E_e = E_e^{\rm max}$ is used for the energy loss in the Sun.
}
\label{Fig4}
\end{figure}
%

In Figure~\ref{Fig4}, we present results for the case when we use the maximum energy loss of boosted $\psi_B$ in the Sun, $\Delta E_B^{\rm Sun}$, by taking the most extreme case $E_e=E_e^{\rm max}$ as explained in Section~\ref{EnergyLoss}.
In this case, the boosted $\psi_B$ loses more kinetic energy, $\thickapprox$ several GeV, than $E_e=E_e^{\rm peak}$ as shown in Figure~\ref{FigDEB}, which is the origin of the deficit of signal events in a low $m_A$ region compared to Figure~\ref{Fig1},
while there is almost no change in a higher $m_A$ region since $E_B^i=m_A \gg \Delta E_B^{\rm Sun}$.
For $\frac{m_B}{\rm GeV} \gtrsim \frac{1}{10}\sqrt{\frac{m_A}{\rm GeV}}$,\footnote{This condition can be easily numerically checked from Eq.~(\ref{Emax}).}
$E_e^{\rm max}$ drops rapidly and $\Delta E_B^{\rm Sun}$ decreases, and consequently the deficit of signals becomes dimmer.

\subsection{Backgrounds}

The dominant backgrounds for the boosted DM signal originate from the charged current interaction of atmospheric neutrinos, i.e., $\nu_e n \to e^- p$.
Super-K has measured the atmospheric neutrino backgrounds for 10.7 years~\cite{Pik:2012qsy}.
In total, 7,755 single-ring zero-decay electron events and 2,105 single-ring electron events have been detected in the energy range of $( 0.1\, {\rm GeV}\, -\, 1.3\, {\rm GeV} )$ and $( 1.33\, {\rm GeV}\, -\, 100\, {\rm GeV} )$, respectively.
For Super-K and Hyper-K, we use all 9,860 events, in the range of $\left( 0.1\, {\rm GeV}\, -\, 100\, {\rm GeV} \right)$ as conservative backgrounds
although higher energy background events are less relevant to $\psi_B$ from a lighter mass of $\psi_A$ which produces a less energetic event.
Thus, we have a yearly background event rate:
\begin{eqnarray}
\frac{N_{\rm BG}}{\Delta T} = 922/{\rm year}\, \left( \frac{V_{\rm exp}}{2.24\times 10^4\, {\rm m}^3} \right)\,,
\end{eqnarray}
where $V_{\rm exp}$ is the volume of the detector.
For PINGU, we use all the events including multi-ring and $\mu$-like events in the $\left( 1.33\, {\rm GeV}\, -\, 100\, {\rm GeV} \right)$ energy range due to a higher $E_e^{\rm th}$ and a poor reconstruction efficiency of the Cherenkov rings~\cite{Agashe:2014yua}.
After rescaling by the effective detector volume of PINGU, $5\times 10^5\, {\rm m}^3$, we obtain a background rate of 14,100/year.

The boosted DM flux comes from a point-like source, the Sun, while the atmospheric neutrino backgrounds are almost uniform in the entire sky.
Thus, the background reduction is governed by the angular resolution of each experiment:
\begin{eqnarray}
N_{\rm BG}^{\theta_{\rm res}} = \frac{1 - \cos\theta_{\rm res}}{2}\, N_{\rm BG}\,.
\end{eqnarray}
For experiments that we consider, a yearly background event rate is reduced as follows.
\begin{eqnarray}
{\rm Super-K}:~ \frac{N_{\rm BG}^{3^\circ}}{\Delta T} &=& 0.63/{\rm year}\,,\\
{\rm Hyper-K}:~ \frac{N_{\rm BG}^{3^\circ}}{\Delta T} &=& 15.8/{\rm year}\,,\\
{\rm PINGU}:~ \frac{N_{\rm BG}^{23^\circ}}{\Delta T} &=& 562/{\rm year}\,.
\end{eqnarray}

\subsection{Detection prospects}

%
\begin{figure}[t]
\begin{center}
\includegraphics[width=0.40\linewidth]{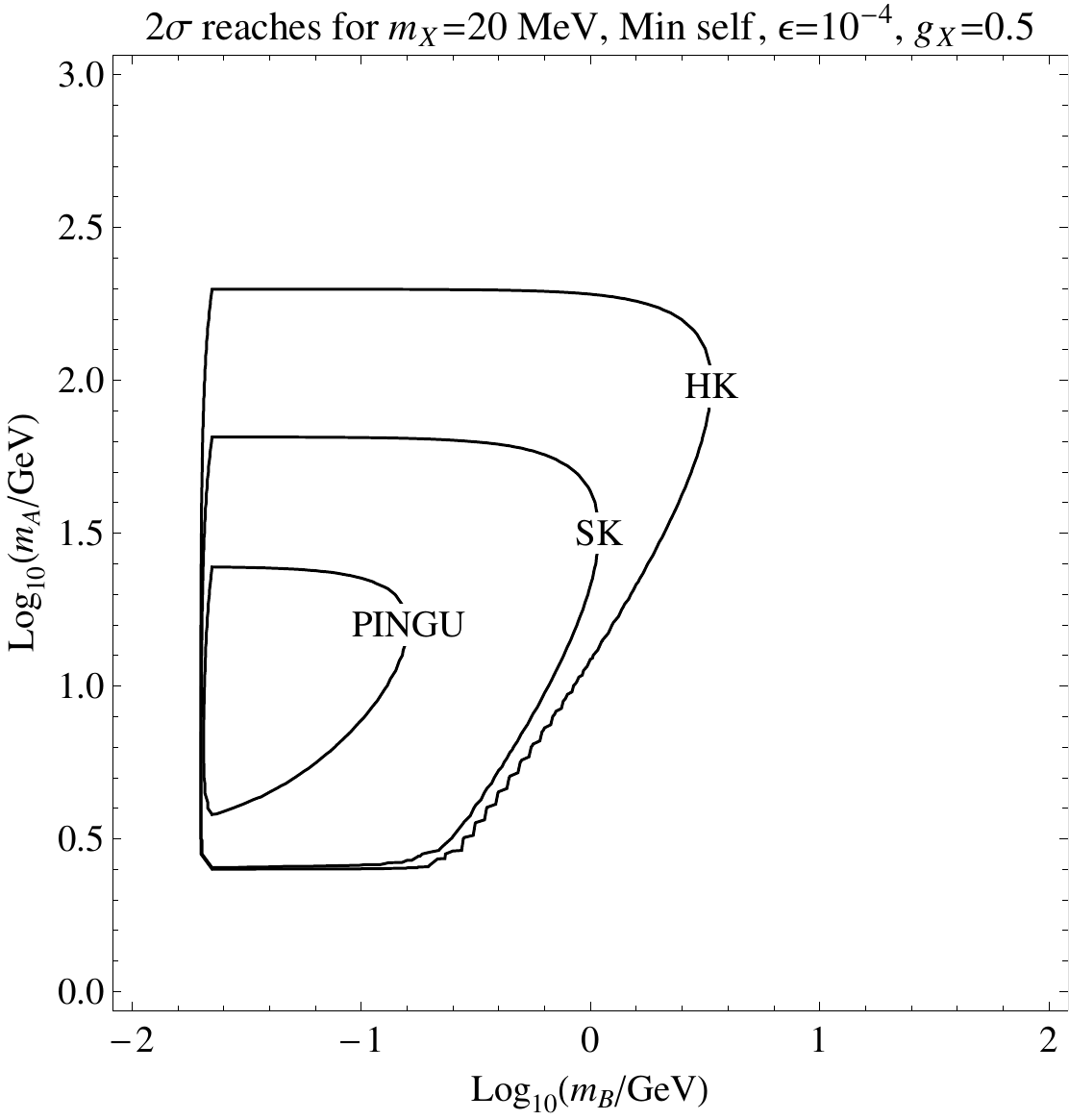}
\hspace*{0.7cm}
\includegraphics[width=0.40\linewidth]{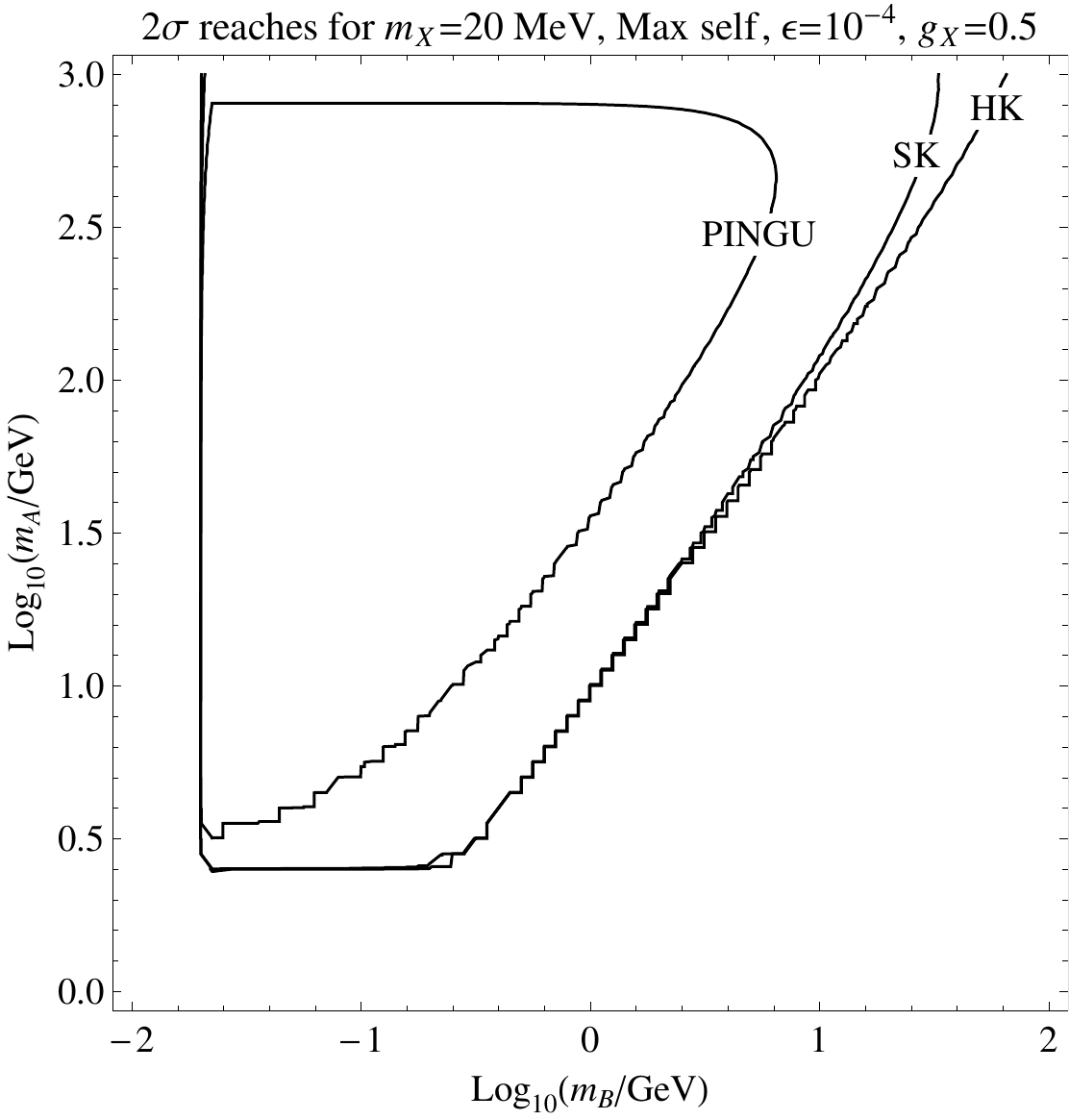}
\vspace*{0.4cm}

\includegraphics[width=0.40\linewidth]{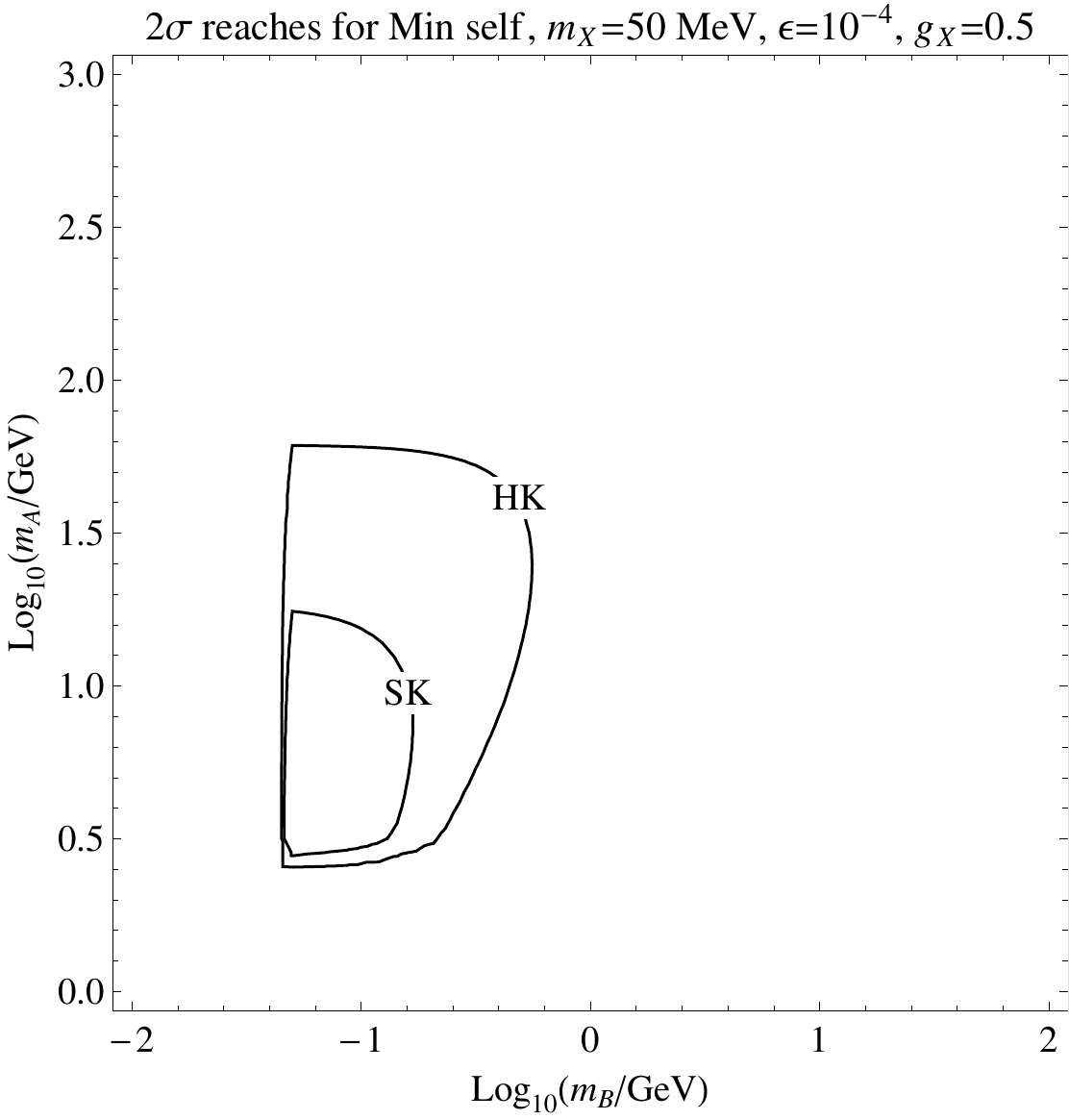}
\hspace*{0.7cm}
\includegraphics[width=0.40\linewidth]{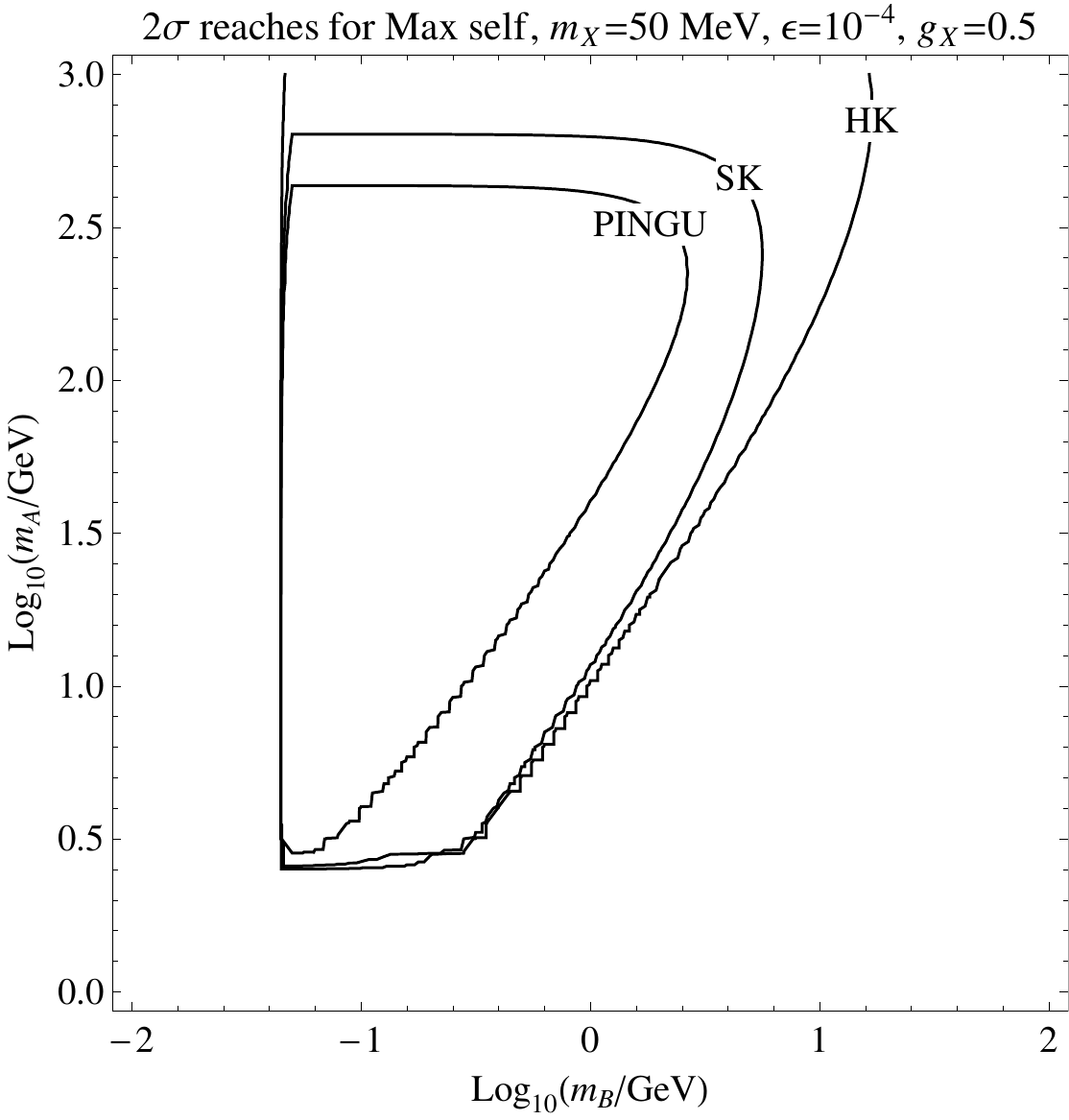}
\end{center}
\vspace*{-0.7cm}
\caption{2$\sigma$ sensitivities with 10.7 years of data
at Super-K (SK), Hyper-K (HK), and PINGU in the ($m_B$, $m_A$) plane
for  $m_X=20$ MeV (top) and $50$ MeV (bottom), and
Min (left) and Max (right) self-interaction, respectively.
}
\label{Fig5}
\end{figure}
%

In Figure~\ref{Fig5}, we present sensitivity curves at the $2 \sigma$ level for the 10.7 years of running period with the definition of the signal significance ($S$),
\begin{eqnarray}
S^{\theta_{\rm res}} \equiv \frac{N_{\rm sig}}
{\sqrt{N^{\theta_{\rm res}}_{\rm BG}}}\,,
\end{eqnarray}
where $\theta_{\rm res}$ is the angular resolution of each experiment in the Cherenkov-emitted electron direction.
Note that the number of signal events, $N_{\rm sig}$, is independent of ${\theta_{\rm res}}$.
While the $2 \sigma$ sensitivity curve for Super-K is obtained based on the 10.7 years of Super-K data, those for Hyper-K and PINGU are estimated assuming the same running time for the convenience of comparison.
PINGU can detect more signal events than Super-K as shown in Figures \ref{Fig1} and \ref{Fig3}, but will be able to cover a smaller parameter region due to its poorer angular resolution.
Thus, Hyper-K is currently the best experiment to find the boosted DM flux from the Sun due to its larger volume, even though it has the same angular resolution as Super-K.

\subsection{Detection of boosted dark matter from the Earth}\label{Earth}

Finally we would like to comment on the detection prospects of boosted DM $\psi_B$ from the Earth.
As shown in Section 9.4 of Ref.~\cite{Jungman:1995df}, the ratio between the capture rates in the Sun and Earth is given by
\begin{eqnarray}
C_c^{\rm Earth}/C_c^{\rm Sun} \approx 10^{-9} \, ,
\end{eqnarray}
due to the much smaller mass ($M_\oplus/M_\odot \approx 3\times 10^{-6}$) and escape velocity ($v_{\rm esc}^\oplus/v_{\rm esc}^\odot \approx 10^{-2}$) of the Earth.
In the case of no self-interaction and no evaporation, the boosted DM flux is simply proportional to $C_c / R^2$ where R is the distance between a detector and a source of the boosted DM.
For the Earth, very tiny capture rates $C_c^{\rm Earth}$ can be compensated by the smaller distance from the source of the boosted DM.
In the absence of self-interaction and evaporation, we find
\begin{eqnarray}
\frac{\Phi_B^{\rm Earth}}{\Phi_B^{\rm Sun}} \approx \frac{C_c^{\rm Earth}}{C_c^{\rm Sun}}\, \frac{R_{\rm Sun}^2}{R_\oplus^2} \approx 0.5\,,
\end{eqnarray}
where $R_{\rm Sun} \simeq 1.5 \times 10^8$ km is the distance between the Sun and the Earth and $R_\oplus \simeq 6.4 \times 10^3$ km is the radius of the Earth.
The evaporation effect is efficient only for a very low DM mass $m_{\rm DM} < 3-4$ GeV for the Sun, whereas it is important up to $m_{\rm DM} \lesssim 12$ GeV for the Earth due to the much smaller escape velocity of the Earth \cite{Griest:1986yu, Gould:1987ju, Gould:1987ir}.

The DM self-capture rate $C_s$ is proportional to $N_\chi$ (see Eq.~(\ref{SelfCapture})), and the seed of $N_\chi$ is determined by $C_c$ since $N_\chi(0)=0$.
A smaller $C_c$ therefore induces a smaller $C_s$, and
consequently, self-capture is negligible for the Earth \cite{Zentner:2009is} since $C_c^{\rm Earth}/C_c^{\rm Sun} \approx 10^{-9}$.
In summary, without self-interactions, $\Phi_B^{\rm Earth}$ could be comparable to $\sim 0.5 \, \Phi_B^{\rm Sun}$ or less depending on $m_{\rm DM}$.
However, with self-interactions, $\Phi_B^{\rm Earth}$ is much smaller than $\Phi_B^{\rm Sun}$ since self-capture is negligible for the Earth,
while it enhances the flux significantly in the Sun.

Moreover the Earth is not a point-like source due to the short distance from the source unlike the Sun.
Even if we consider only the inner core of the Earth, $R_{\rm in-core}^{\rm Earth} \approx 1.2 \times 10^3$ km, we should integrate over a $\sim 34^\circ$ cone around the center of the Earth.
Thus, the background events by atmospheric neutrinos are governed by $\theta \simeq 34^\circ$ instead of the angular resolution of each experiment $\theta_{\rm res}$.
Especially for Super-K and Hyper-K with $\theta_{\rm res}\simeq 3^\circ$, we have $\sim 125$ times larger backgrounds events for the boosted DM flux from the Earth compared to the analysis for the Sun.
Even for the case of $\Phi_B^{\rm Earth} \approx \Phi_B^{\rm Sun}$, the final signal significance $S$ for the Earth signals is therefore much less than that for the solar signals.

\section{Conclusion}

The current paradigm of CDM is extremely successful in explaining much of cosmological and astronomical data.
However there still exist several questions including the missing satellite problem,
the core/cusp problem, issues in substructure, and the too-big-to-fail problem.
Self-interacting multi-component dark matter provides profound insight into these problems and
hence it is imperative to consider detection prospects of such scenarios.

In this paper, we have investigated the discovery potential of the boosted DM flux from the Sun in large volume neutrino experiments such as Super-K, Hyper-K, and PINGU.
We considered a model in Ref. \cite{Agashe:2014yua} and
additionally introduced self-interaction to the heavier secluded DM.
The heavier DM particle $\psi_A$ is thermalized with the assistance of the other DM $\psi_B$, i.e. through the {\it assisted freeze-out} mechanism~\cite{Belanger:2011ww}.
The $\psi_A$ can be effectively captured in the Sun due to its self-interaction in the range favored by cosmological simulations and measurements.
The accumulated dark matter then annihilates to the lighter sister with a large Lorentz boost.
In a large volume neutrino detector, the boosted dark matter can be probed by measuring electrons via
elastic scattering with the $\psi_B$ from the Sun.
Certainly a detector with a larger volume and lower electron energy threshold  $E_e^{\rm th}$ performs better.
Improvement on angular resolution at future experiments is crucial in detecting boosted dark matter from the Sun,
as atmospheric neutrino backgrounds do not exhibit any directionality and can be efficiently reduced for a point-like source, the Sun, in contrast to the the galactic center region.

Searches for the signals by the boosted DM fluxes from the Sun (this study) and the GC (Ref.~\cite{Agashe:2014yua}) are complementary,
since the parameter space that is accessible to one is not accessible to the other.
There are two main complementary effects.
First, for the GC signal the value of $\epsilon$ needs to be higher, $\epsilon \sim 10^{-3}$, to overcome the small flux, while for the solar signal $\epsilon$ needs to be smaller, $\epsilon \sim 10^{-4}$, to reduce the energy loss inside the Sun.
Second, without DM self-interactions one prefers an NFW-like cusp DM halo profiles to enhance the GC signal, whereas with DM self-interactions a more cored profile to relatively suppress the GC signal and allow the solar signal.
Thus, searches for both signals from the GC and the Sun should be conducted at the same time.

\vspace{0.5 cm}
{\bf Note added:} Near completion of this study, we have noticed that a related study, Ref. \cite{Berger:2014sqa} was submitted to {\tt arXiv.org}.
We have used a different reference model and focused on implications of self-interaction in the context of the boosted dark matter from the Sun.

\vspace{-0.3cm}
\begin{acknowledgements}
\vspace{-0.3cm}
We would like to thank Yen-Hsun Lin for providing valuable information on the calculation of the DM evolution, and Chang Hyon Ha and Jesse Thaler for helpful discussions.
K.K. is supported by the U.S. DOE under Grant No. DE-FG02-12ER41809 and by the University of Kansas General Research Fund allocation 2301566.
J.C.P. is supported by the Basic Science Research Program through the National Research Foundation of Korea funded by the Ministry of Education (NRF-2013R1A1A2061561). G.M is supported in part by the National Research Foundation of South Africa under Grant No. 88614.
\end{acknowledgements}


\begin{thebibliography}{99}




\bibitem{Arrenberg:2013rzp}
  S.~Arrenberg, H.~Baer, V.~Barger, L.~Baudis, D.~Bauer, J.~Buckley, M.~Cahill-Rowley and R.~Cotta {\it et al.},
  arXiv:1310.8621 [hep-ph].


\bibitem{Konar:2009qr}
  P.~Konar, K.~Kong, K.~T.~Matchev and M.~Park,
  JHEP {\bf 1004}, 086 (2010)
  [arXiv:0911.4126 [hep-ph]].



\bibitem{Moore:1994yx}
  B.~Moore,
  Nature {\bf 370}, 629 (1994);
  R.~A.~Flores and J.~R.~Primack,
  Astrophys.\ J.\  {\bf 427}, L1 (1994)
  [astro-ph/9402004];
  J.~F.~Navarro, C.~S.~Frenk and S.~D.~M.~White,
  Astrophys.\ J.\  {\bf 490}, 493 (1997)
  [astro-ph/9611107].


\bibitem{Walker:2011zu}
  M.~G.~Walker and J.~Penarrubia,
  Astrophys.\ J.\  {\bf 742}, 20 (2011)
  [arXiv:1108.2404 [astro-ph.CO]];
  S.~H.~Oh, W.~J.~G.~de Blok, E.~Brinks, F.~Walter and R.~C.~Kennicutt, Jr,
  Astron.\ J.\  {\bf 141}, 193 (2011)
  [arXiv:1011.0899 [astro-ph.CO]];
  S.~H.~Oh, W.~J.~G.~de Blok, E.~Brinks, F.~Walter and R.~C.~Kennicutt, Jr,
  Astron.\ J.\  {\bf 141}, 193 (2011)
  [arXiv:1011.0899 [astro-ph.CO]].


\bibitem{BoylanKolchin:2011de}
  M.~Boylan-Kolchin, J.~S.~Bullock and M.~Kaplinghat,
  Mon.\ Not.\ Roy.\ Astron.\ Soc.\  {\bf 415}, L40 (2011)
  [arXiv:1103.0007 [astro-ph.CO]];
  M.~Boylan-Kolchin, J.~S.~Bullock and M.~Kaplinghat,
  Mon.\ Not.\ Roy.\ Astron.\ Soc.\  {\bf 422}, 1203 (2012)
  [arXiv:1111.2048 [astro-ph.CO]].


\bibitem{Lovell:2013ola}
  M.~R.~Lovell, C.~S.~Frenk, V.~R.~Eke, A.~Jenkins, L.~Gao and T.~Theuns,
  Mon.\ Not.\ Roy.\ Astron.\ Soc.\  {\bf 439}, 300 (2014)
  [arXiv:1308.1399 [astro-ph.CO]].


\bibitem{Spergel:1999mh}
  D.~N.~Spergel and P.~J.~Steinhardt,
  Phys.\ Rev.\ Lett.\  {\bf 84}, 3760 (2000)
  [astro-ph/9909386].


\bibitem{Rocha:2012jg}
  M.~Rocha, A.~H.~G.~Peter, J.~S.~Bullock, M.~Kaplinghat, S.~Garrison-Kimmel, J.~Onorbe and L.~A.~Moustakas,
  Mon.\ Not.\ Roy.\ Astron.\ Soc.\  {\bf 430}, 81 (2013)
  [arXiv:1208.3025 [astro-ph.CO]];
  A.~H.~G.~Peter, M.~Rocha, J.~S.~Bullock and M.~Kaplinghat,
  arXiv:1208.3026 [astro-ph.CO].

\bibitem{Randall:2007ph}
  S.~W.~Randall, M.~Markevitch, D.~Clowe, A.~H.~Gonzalez and M.~Bradac,
  Astrophys.\ J.\  {\bf 679}, 1173 (2008)
  [arXiv:0704.0261 [astro-ph]].

\bibitem{Zavala:2012us}
  J.~Zavala, M.~Vogelsberger and M.~G.~Walker,
  Monthly Notices of the Royal Astronomical Society: Letters {\bf 431}, L20 (2013)
  [arXiv:1211.6426 [astro-ph.CO]].



\bibitem{D'Eramo:2010ep}
  F.~D'Eramo and J.~Thaler,
  JHEP {\bf 1006}, 109 (2010)
  [arXiv:1003.5912 [hep-ph]].

\bibitem{Belanger:2012vp}
  G.~Belanger, K.~Kannike, A.~Pukhov and M.~Raidal,
  JCAP {\bf 1204}, 010 (2012)
  [arXiv:1202.2962 [hep-ph]].

\bibitem{Belanger:2011ww}
  G.~Belanger and J.~C.~Park,
  JCAP {\bf 1203}, 038 (2012)
  [arXiv:1112.4491 [hep-ph]].



\bibitem{Huang:2013xfa}
  J.~Huang and Y.~Zhao,
  JHEP {\bf 1402}, 077 (2014)
  [arXiv:1312.0011 [hep-ph]].

\bibitem{Agashe:2014yua}
   K.~Agashe, Y.~Cui, L.~Necib and J.~Thaler,
  JCAP {\bf 1410}, no. 10, 062 (2014)
  [arXiv:1405.7370 [hep-ph]].

\bibitem{Berger:2014sqa}
  J.~Berger, Y.~Cui and Y.~Zhao,
  arXiv:1410.2246 [hep-ph].




\bibitem{Okun:1982xi}
  L.~B.~Okun,
  Sov.\ Phys.\ JETP {\bf 56}, 502 (1982)
  [Zh.\ Eksp.\ Teor.\ Fiz.\  {\bf 83}, 892 (1982)];
  B.~Holdom,
  Phys.\ Lett.\ B {\bf 166}, 196 (1986).
  J.~H.~Huh, J.~E.~Kim, J.~C.~Park and S.~C.~Park,
  Phys.\ Rev.\ D {\bf 77}, 123503 (2008)
  [arXiv:0711.3528 [astro-ph]];
  E.~J.~Chun and J.~C.~Park,
  JCAP {\bf 0902}, 026 (2009)
  [arXiv:0812.0308 [hep-ph]];
  J.~C.~Park and S.~C.~Park,
  Phys.\ Lett.\ B {\bf 718}, 1401 (2013)
  [arXiv:1207.4981 [hep-ph]];
  G.~Belanger, A.~Goudelis, J.~C.~Park and A.~Pukhov,
  JCAP {\bf 1402}, 020 (2014)
  [arXiv:1311.0022 [hep-ph]].

\bibitem{Chun:2010ve}
  E.~J.~Chun, J.~C.~Park and S.~Scopel,
  JHEP {\bf 1102}, 100 (2011)
  [arXiv:1011.3300 [hep-ph]].


\bibitem{Essig:2013lka}
  R.~Essig, J.~A.~Jaros, W.~Wester, P.~H.~Adrian, S.~Andreas, T.~Averett, O.~Baker and B.~Batell {\it et al.},
  arXiv:1311.0029 [hep-ph].


\bibitem{Bandyopadhyay:2011qm}
  P.~Bandyopadhyay, E.~J.~Chun and J.~C.~Park,
  JHEP {\bf 1106}, 129 (2011)
  [arXiv:1105.1652 [hep-ph]].





\bibitem{Steigman:1997vs}
  G.~Steigman, C.~L.~Sarazin, H.~Quintana and J.~Faulkner,
  Astron.\ J.\  {\bf 83}, 1050 (1978);
  D.~N.~Spergel and W.~H.~Press,
  Astrophys.\ J.\  {\bf 294}, 663 (1985);
  W.~H.~Press and D.~N.~Spergel,
  Astrophys.\ J.\  {\bf 296}, 679 (1985);
  J.~Faulkner and R.~L.~Gilliland,
  Astrophys.\ J.\  {\bf 299}, 994 (1985).

\bibitem{Griest:1986yu}
  K.~Griest and D.~Seckel,
  Nucl.\ Phys.\ B {\bf 283}, 681 (1987)
  [Erratum-ibid.\ B {\bf 296}, 1034 (1988)].

\bibitem{Gould:1987ju}
  A.~Gould,
  Astrophys.\ J.\  {\bf 321}, 560 (1987).

\bibitem{Busoni:2013kaa}
  G.~Busoni, A.~De Simone and W.~C.~Huang,
  JCAP {\bf 1307}, 010 (2013)
  [arXiv:1305.1817 [hep-ph]].

\bibitem{Zentner:2009is}
  A.~R.~Zentner,
  Phys.\ Rev.\ D {\bf 80}, 063501 (2009)
  [arXiv:0907.3448 [astro-ph.HE]].

\bibitem{Albuquerque:2013xna}
  I.~F.~M.~Albuquerque, C.~Perez de Los Heros and D.~S.~Robertson,
  JCAP {\bf 1402}, 047 (2014)
  [arXiv:1312.0797 [astro-ph.CO]].

\bibitem{Chen:2014oaa}
  C.~S.~Chen, F.~F.~Lee, G.~L.~Lin and Y.~H.~Lin,
  JCAP {\bf 1410}, no. 10, 049 (2014)
  [arXiv:1408.5471 [hep-ph]].


\bibitem{SolarModelFile}
http://www.sns.ias.edu/$\sim$jnb/SNdata/sndata.html


\bibitem{Baratella:2013fya}
  P.~Baratella, M.~Cirelli, A.~Hektor, J.~Pata, M.~Piibeleht and A.~Strumia,
  JCAP {\bf 1403}, 053 (2014)
  [arXiv:1312.6408 [hep-ph]].


\bibitem{Kappl:2011kz}
  R.~Kappl and M.~W.~Winkler,
  Nucl.\ Phys.\ B {\bf 850}, 505 (2011)
  [arXiv:1104.0679 [hep-ph]].


\bibitem{Gould:1991hx}
  A.~Gould,
  Astrophys.\ J.\  {\bf 388}, 338 (1992).


\bibitem{Agashe:2014kda}
  K.~A.~Olive {\it et al.}  [Particle Data Group Collaboration],
  Chin.\ Phys.\ C {\bf 38}, 090001 (2014), Section 32 (Passage of particle through matter).


\bibitem{Jungman:1995df}
  G.~Jungman, M.~Kamionkowski and K.~Griest,
  Phys.\ Rept.\  {\bf 267}, 195 (1996)
  [hep-ph/9506380].

\bibitem{Gould:1987ir}
  A.~Gould,
  Astrophys.\ J.\  {\bf 321}, 571 (1987).

\bibitem{Zentner:2009is}
  A.~R.~Zentner,
  Phys.\ Rev.\ D {\bf 80}, 063501 (2009)
  [arXiv:0907.3448 [astro-ph.HE]].


\bibitem{Pik:2012qsy}
  L.~K.~Pik,
  PhD dissertation, University of Tokyo, 2012.


\end{thebibliography}
\end{document}